\documentclass[10pt,journal]{IEEEtran}

\usepackage{amsfonts}
\usepackage{extarrows}
\usepackage{enumerate}
\usepackage{algorithm,algorithmic}
\usepackage{graphicx}
\usepackage{color}
\usepackage{amsmath,amssymb,amsthm,mathrsfs,amsfonts,dsfont}
\usepackage[square,comma,sort&compress,numbers]{natbib}
\usepackage[colorlinks=true,linkcolor=black,citecolor=black]{hyperref}
\usepackage{cases}
\usepackage{empheq}
\usepackage{subcaption}
\newtheorem{itassumption}{Assumption}

\newtheorem{itremark}{Remark}

\allowdisplaybreaks[4]

\usepackage[font=small]{caption}

\begin{document}

\title{Quantized Consensus under Data-Rate Constraints and DoS Attacks: A Zooming-In and Holding Approach}

\author{{Maopeng Ran, Shuai Feng, Juncheng Li, and Lihua Xie, \emph{Fellow, IEEE}}
\thanks{M. Ran is with the School of Automation Science and Electrical Engineering, Beihang University, Beijing 100191, China (email: ranmp@buaa.edu.cn).}
 \thanks{J. Li, and L. Xie (corresponding author) are with the School of Electrical and Electronic Engineering, Nanyang Technological University, Singapore 639798 (email: juncheng001@e.ntu.edu.sg; elhxie@ntu.edu.sg). }
 \thanks{S. Feng is with the School of Automation, Nanjing University of Science and Technology, Nanjing 210094, China (email: s.feng@njust.edu.cn).}}

\maketitle

\begin{abstract}

This paper is concerned with the quantized consensus problem for uncertain nonlinear multi-agent systems under data-rate constraints and Denial-of-Service (DoS) attacks. The agents are modeled in strict-feedback form with unknown nonlinear dynamics and external disturbance. Extended state observers (ESOs) are leveraged to estimate agents' total uncertainties along with their states. To mitigate the effects of DoS attacks, a novel dynamic quantization with zooming-in and holding capabilities is proposed. The idea is to zoom-in and hold the variable to be quantized if the system is in the absence  and presence of DoS attacks, respectively. The control protocol is given in terms of the outputs of the ESOs and the dynamic-quantization-based encoders and decoders. We show that, for a connected undirected network, the developed control protocol is capable of handling any DoS attacks inducing bounded consecutive packet losses with merely 3-level quantization. The application of the zooming-in and holding approach to known linear multi-agent systems is also discussed.

\end{abstract}


\IEEEpeerreviewmaketitle

\section{Introduction}


\IEEEPARstart{C}ONSENSUS over digital networks is known as one of the most important issues in cooperative control of multi-agent systems \cite{Oh-2015}. The success of consensus relies on efficient and reliable information exchange between the agents. According to the basic communication principles, the capacity of a real digital network is limited, i.e., at each time step or time interval, only limited data can be transmitted reliably between the agents. On the other hand, the digital network is vulnerable to malicious cyber attacks, such as Denial-of-Service (DoS) \cite{Sirdhar-2012}. DoS attacks refer to destroying the information availability, which may deteriorate the consensus performance and even lead to instability. Thus, it is of theoretical significance and practical relevance to investigate the consensus problem for multi-agent systems under data-rate constraints and DoS attacks.

\subsection{Literature Review}

Data-rate constraints for multi-agent systems can be modeled by introducing a quantization-encoding-decoding process with a finite number of quantization levels \cite{Li-2011}. At each time step, the sender encodes the quantized information and sends out the code via the digital network. When the neighbors receive the code, they use a decoding algorithm to reconstruct the information. Primary studies on the quantized consensus problem have been on first-order integrator multi-agent systems \cite{Kash-2007,Aysal-2008,Carli-2010,Li-2011,Lin-2013}. In \cite{Kash-2007}, the states of the agents were assumed to be integer-valued. In \cite{Aysal-2008}, a random dither was added to the sensor state before quantization  to make the quantization error a ``white'' noise. In  \cite{Li-2011}, with the use of a finite-level uniform quantizer and an exponentially decaying scaling function, it was shown that the average consensus can be achieved with one-bit data rate for a connected undirected network. The idea in \cite{Li-2011} was further generalized to multi-agent systems with higher-order dynamics \cite{Qiu-2016,You-2011,Meng-2017,Dong-2019,Ran-2019}.

Recent years have also witnessed a growing interest towards control systems under DoS attacks. In \cite{Persis-2015}, DoS attacks were characterized by their levels of average frequency and duration under which input-to-state stability of the closed-loop system can be preserved. Such a characterization of DoS attacks has motivated many interesting results, e.g., \cite{Persis-2016,Feng-2017,Persis-2018,Yang-2018}. For multi-agent systems under DoS attacks, Deng and Wen \cite{Deng-2020} developed a resilient fault-tolerant approach to handle the secure cooperative problem in the presence of actuator faults and DoS attacks. In \cite{Hu-2020} and \cite{Tang-2021}, the event-triggered strategy was introduced to linear and nonlinear multi-agent systems under DoS attacks, respectively. We mention that the communication data-rate was implicitly assumed to be infinite in \cite{Deng-2020,Hu-2020,Tang-2021}, since the agents can communicate exact information with their neighbors.

For the simultaneous consideration of data-rate constraints and DoS attacks,  a few recent works have been done for centralized systems \cite{Waka-2020,Feng-2021,Guan-2021,Liu-2021,Liu-2021b}, and multi-agent systems \cite{Feng-2020b}. In \cite{Waka-2020}, a dynamic encoding scheme was developed, and sufficient conditions on DoS frequency and duration for exponential state convergence were obtained. In \cite{Feng-2021,Guan-2021}, the trade-off between system resilience against DoS attacks and data-rate was analyzed, in which the data-rate conditions are dependent on the plant state matrix, and DoS frequency and duration. In \cite{Liu-2021b}, the input-to-state stability (ISS) of linear systems with quantized state measurements under external disturbances and DoS attacks was studied. In \cite{Liu-2021}, the scenario that both input (controller-to-plant) and output (plant-to-controller) channels are subject to data-rate constraints and DoS attacks was investigated. An interesting property of the approach in  \cite{Liu-2021} is that when only the output channel is subject to DoS attacks, the widely adopted network acknowledgement signal is not required.  For multi-agent systems under data-rate constraints and DoS attacks, Feng and Ishii \cite{Feng-2020b} proposed a dynamic quantization approach with \emph{zooming-in and zooming-out} capabilities. Specifically, the scaling function used for dynamic quantization zooms-in and out the variable to be quantized in the absence and presence of DoS, respectively. It should be pointed out that the required data-rate for each paired agents in \cite{Feng-2020b} increases to infinity if the number of the agents $N\rightarrow \infty$. What is more, the tolerable level of DoS attacks specified by their average frequency and duration in \cite{Feng-2020b} may be low, since the quantized control scheme needs to zoom-out under DoS for compensating the diverging consensus error.

By investigation of the existing literature, there are two important issues that are noteworthy. On one hand, it still lacks a solution to the quantized consensus problem of general \emph{uncertain nonlinear} multi-agent systems under data-rate constraints and DoS attacks. The approaches in \cite{Waka-2020,Feng-2021,Guan-2021,Liu-2021,Liu-2021b,Feng-2020b} are for certain linear systems. However, most practical control systems are inherently uncertain and nonlinear \cite{Khalil-2002,Isi-1989}. On the other hand, it is highly desirable to develop new techniques to achieve lower data-rate and higher tolerance to DoS attacks. 

\subsection{Main Contribution}

The theme of this paper is to investigate the quantized consensus problem for uncertain nonlinear multi-agent systems under data-rate constraints and DoS attacks. The agents are modeled in strict-feedback form with unknown nonlinear dynamics and external disturbance. When DoS occurs, the information availability between the agents can be destroyed,  and the consensus becomes difficult under DoS attacks. Therefore, an important technical difficulty we need to overcome is to efficiently govern the multi-agent systems during the attack periods, with the uncertain nonlinear agent dynamics well-handled.  Based on the intuitive consideration that it is expected to lead to a higher tolerance to DoS attacks if the multi-agent system can hold its status instead of having a diverging trend  when it is attacked, we develop a new dynamic quantization scheme  which we refer to as \emph{zooming-in and holding}. The new scheme is significantly different from that in the state-of-the-art result \cite{Feng-2020b}, as it zooms-in the variable to be quantized in the absence of DoS attacks, and holds the variable in the presence of DoS attacks. What is more, we employ an extended state observer (ESO) \cite{Khalil-2017,Han-2009} to  estimate the system unknown nonlinear dynamics and external disturbance, and compensate them in real time in the control action.
\emph{In summary, the main contribution of this paper is the development of an ESO-based dynamic encoding-decoding scheme with zooming-in and holding capabilities for uncertain nonlinear multi-agent systems under data-rate constraints and DoS attacks.}

Compared with the existing literature, the advantages of the control scheme developed in this paper include:
\begin{enumerate}
\item It is a more practical solution to the quantized consensus problem under data-rate constraints and DoS attacks. The developed scheme is capable of handling uncertain nonlinear dynamics and external disturbance. More importantly, it is tolerable to serious DoS attacks. We only require that the consecutive packet losses induced by DoS attacks be bounded. This is a minimum requirement since a multi-agent system becomes $N$ single systems if an attack is always active.
\item It is capable of leading to better consensus performance. This is due to the nature of the zooming-in and holding mechanism, in which the consensus error keeps invariant during DoS attacks. On the contrary, the consensus error may diverge during DoS attacks for the zooming-in and zooming-out mechanism \cite{Feng-2020b}.
\item It achieves theoretically the lowest data-rate. We show that merely 3-level quantization suffices to guarantee quantized consensus, which is independent from the number of the agents and DoS attacks.
\end{enumerate}

The remainder of this paper is organized as follows. Section \ref{Sec_prob} provides some preliminaries and problem formulation. The proposed control scheme for uncertain nonlinear multi-agent systems is presented in Section \ref{Sec_non}. In Section \ref{Sec_lin}, for ease of comparison, we apply the zooming-in and holding approach to known linear multi-agent systems. Two examples are provided in Section \ref{Sec_exp}. Finally, Section \ref{Sec_con} concludes the paper.

\section{Preliminaries and Problem Formulation}\label{Sec_prob}

\subsection{Notation}

In this paper, we use $\mathbb{R}$, $\mathbb{R}^n$, and $\mathbb{R}^{n\times m}$ to denote the set of real numbers, $n$-dimensional real column vectors, and $n\times m$ real matrices, respectively. For a given vector or matrix $A$, its transpose, Euclidean norm, and $\infty$-norm are denoted by $A^{\textrm{T}}$, $\|A\|$, and $\|A\|_{\infty}$, respectively. For a given positive number $\nu$, the maximum integer less than or equal to $\nu$ is denoted by $\lfloor \nu  \rfloor$; the minimum integer greater than or equal to $\nu$ is denoted by $\lceil \nu \rceil$. $I_n$ denotes the identity matrix of  dimension $n$. $\textbf{1}_n$ and $\textbf{0}_n$ denote the $n$-dimensional column vector with all ones and zeros, repectively. Let $J_n=(1/n)\textbf{1}_n\textbf{1}^{\textrm{T}}_n$.  Big $O$-notation in terms of $\nu$ is denoted as $O(\nu)$. Let $\textrm{sat}(\cdot)$ be the unity saturation function defined by $\textrm{sat}(\nu)=\textrm{sign}(\nu)\cdot \min\{1,|\nu|\}$.

\subsection{Graph Theory}

For a group of $N$ agents, the corresponding graph $\mathcal{G}$ can be expressed by a triple $\{\mathcal{V},\mathcal{E}, \mathcal{A}\}$, where $\mathcal{V}=\{1, 2, \ldots, N\}$ is the set of nodes, $\mathcal{E}\subset\mathcal{V}\times\mathcal{V}$  is the set of edges, and $\mathcal{A}=[a_{ij}]\in \mathbb{R}^{N\times N}$ is the adjacency matrix with $a_{ij}=1$ if $(i,j)\in \mathcal{E}$, and $a_{ij}=0$ otherwise. It is assumed $a_{ii}=0$, i.e., no self-loop in $\mathcal{G}$. The neighborhood of agent $i$ is denoted by $\mathcal{N}_i=\{j\in\mathcal{V}|(i,j)\in \mathcal{E}\}$. A sequence of edges $(i_1,i_2)$, $(i_2,i_3)$, $\ldots$, $(i_{j-1}, i_j)$ is called a  path from agent $i_1$ to agent $i_j$. If for any two agents $i, j\in\mathcal{V}$, there exists a  path from agent $i$ to agent $j$,  $\mathcal{G}$ is called a connected graph. The in-degree of agent $i$ is represented by $\textrm{deg}_i=\sum_{j=1}^{N}a_{ij}$, and $d^*=\max_{1\leq i\leq N}\textrm{deg}_i$ is the degree of $\mathcal{G}$. Denote $\mathcal{D}=\textrm{diag}\{\textrm{deg}_1,\ldots, \textrm{deg}_N\}$ and the Lpalacian matrix of $\mathcal{G}$ by $\mathcal{L}=\mathcal{D}-\mathcal{A}$. If $\mathcal{A}$ is a symmetric matrix, $\mathcal{G}$ is called an undirected graph, and the eigenvalues of the Laplacian matrix $\mathcal{A}$ in an ascending order are denoted by $0=\lambda_1< \lambda_2\leq\cdots \leq \lambda_N$ \cite{Fie-1973}.

\subsection{Problem Formulation}

Consider a multi-agent system consisting of $N$ agents described by the following single-input-single-output uncertain nonlinear dynamics:
\begin{equation}\label{eq1}
 \left\{
  \begin{aligned}
          \dot{x}_{ij}= & f_{ij}(x_{i1}, \ldots, x_{ij})+x_{i,j+1}, ~1\leq j\leq r-1, \\
          \dot{x}_{ir}=& f_{ir}(x_i, z_i, \omega_i)+u_i, \\
          \dot{z}_i  =& f_{i0}(x_i,z_i,\omega_i), \\
                y_i=& x_{i1}, ~i=1, \ldots, N,
        \end{aligned} \right.
\end{equation}
where $x_i=[x_{i1}, x_{i2}, \ldots, x_{ir}]^{\rm{T}}\in\mathbb{R}^{r}$ is the agent state, $z_i\in\mathbb{R}^{n_i-r}$ is the state of the zero dynamics; $n_i$ is the dimension of agent $i$, $r$ is the relative degree; $y_i\in\mathbb{R}$, $u_i\in\mathbb{R}$, and $\omega_i\in\mathbb{R}^{n_{\omega_i}}$ are the measured output, control input, and external disturbance, respectively; and $f_{i0}\in C^1(\mathbb{R}^{r}\times \mathbb{R}^{n_i-r}\times \mathbb{R}^{n_{\omega_i}},\mathbb{R})$, $f_{ij}\in C^{r+1-j}(\mathbb{R}^j, \mathbb{R})$, $1\leq j\leq r-1$, and $f_{ir}\in C^{1}(\mathbb{R}^{r}\times \mathbb{R}^{n_i-r}\times \mathbb{R}^{n_{\omega_i}},\mathbb{R})$ are unknown locally Lipschitz functions. The strict-feedback system (\ref{eq1}) represents a wide class of physical plants \cite{Khalil-2002}, such as the pendulum with motor dynamics in  \cite{Kwan-1995} and the third-order phase-locked loop in \cite{Harb-2004}.

\emph{Assumption A1:} The communication graph $\mathcal{G}$ among the $N$ agents is undirected and connected.

\emph{Assumption A2:} The external disturbances $\omega_i$ and their derivatives $\dot{\omega}_i$ are bounded.

\emph{Assumption A3:} The internal dynamics $\dot{z}_i=f_{i0}(x_i,z_i,\omega_i)$ are bounded-input-bounded-state (BIBS) stable with respect to  $(x_i,\omega_i)$.


For a real digital network connecting the multi-agent system (\ref{eq1}), information to be transmitted between the agents is to be first quantized and encoded at the transmitter, and then decoded at the receiver. In control protocols to be developed in the sequel, we use the following finite-level uniform quantizer:
\begin{equation}\label{eq20}
 q(\nu)=\left\{
\begin{aligned}
&0, \quad  -1/2<\nu<1/2, \\
&i,  \quad \frac{2i-1}{2}\leq \nu <\frac{2i+1}{2}, ~i=1, 2, \ldots, K-1, \\
&K,  \quad \nu \geq \frac{2K-1}{2}, \\
&-q(-\nu), \quad  \nu \leq -1/2,
\end{aligned} \right.
\end{equation}
where $K\in \{1, 2, \ldots\}$. For a ($2K+1$)-level quantizer specified by (\ref{eq20}), the agent is required to be capable of transmitting $\lceil\textrm{log}_2(2K+1)\rceil$ bits at each time step.
For subsequent use, let $Q(\nu)=[q(\nu_1), \ldots, q(\nu_{\ell})]^{\textrm{T}}\in\mathbb{R}^{\ell}$ with $\nu=[\nu_1, \ldots, \nu_{\ell}]^{\textrm{T}}\in\mathbb{R}^{\ell}$.
On the other hand, it is assumed that the transmission is acknowledgement-based and free of delay. That is, the receivers send acknowledgments to the transmitters immediately if they receive encoded signals successfully \cite{Waka-2020,Guan-2021,Feng-2020b,Liu-2021,Liu-2021b,Feng-2021}. From a communication point of view, this can be realized by requiring that the acknowledgments are sent by a more powerful source.

In this paper,  similar to \cite{Feng-2020b},  we refer to DoS attacks as the event for which all the encoded signals cannot be received by the decoders and it affects all the agents. The attacks are assumed to be launched by an adversarial entity possibly having limited energy. Hence, here we characterize the attacker's action by its capability of inducing consecutive transmission failures by launching DoS.
In addition to DoS, we apply periodic transmission strategy, i.e., the transmission attempts of each agent takes place periodically at $t=kT$, in which $k$ is a non-negative integer and $T>0$ is the sampling interval. Due to DoS attacks, not all the transmission attempts at $kT$ succeed. Then, we let $\{v_\ell T\}=\{v_0 T, v_1 T, v_2T,...\} \subseteq \{kT\}$ with $\ell = 0, 1, ...$ represent the sequence of time instants when the transmissions are successful.

\emph{Assumption A4:} The duration of the interval $[v_\ell T, v_{\ell+1} T)$ calculated as $v_{\ell+1}T - v_\ell T$ with $\ell =  0, 1, ...$ is bounded.

\emph{Remark 1 (DoS Model):} Under DoS attacks, the interval between two consecutive successful transmissions can be prolonged by DoS attacks. Assumption A4 constrains the maximal number of consecutive packet drops induced by DoS attacks.  In \cite{Persis-2015}, a DoS model was developed to describe an attacker's action in terms of the DoS \emph{average frequency and duration}, and the level of DoS was specified by $\frac{1}{T_D}+\frac{T}{\tau_D}$, where $T_D>1$ and $\tau_D>T$ are DoS model parameters. More specifically,  $\frac{1}{T_D}+\frac{T}{\tau_D}=0$ indicates no DoS attacks, and $\frac{1}{T_D}+\frac{T}{\tau_D}=1$  the attack can be always active. Here we mention that the Assumption A4 has some connections with the DoS model  in \cite{Persis-2015}. Based on the model in \cite{Persis-2015}, the papers \cite{Feng-2017, Persis-2018} presented a result concerning the upper bound of the intervals between two consecutive successful transmissions in the presence of DoS attacks. In particular, the value of the upper bound depends on the level of DoS attacks, that is, if $\frac{1}{T_D}+\frac{T}{\tau_D} $ is high, then the upper bound of the intervals between two consecutive successful transmissions can be long. In the context of our paper, this means $v_{\ell+1}T - v_\ell T$ can be large.  What is more, the results of the paper \cite{Feng-2020b} were developed based on the DoS model in \cite{Persis-2015}, in which the upper bound of consecutive packet losses is involved in the design process. Here, since we will design a novel zooming-in and holding quantization strategy, we only require that the number of consecutive packet losses induced by DoS is upper bounded as in Assumption A4, and the exact upper bound is not necessary to be known for design. Our result is also applicable under other DoS models which can imply bounded consecutive packet losses. \IEEEQED

In the following, we aim to solve the quantized consensus problem for uncertain nonlinear multi-agent system (\ref{eq1}) over a digital network under data-rate constraints and DoS attacks. Our main results will be stated in Section \ref{Sec_non}. The special case that the multi-agent system (\ref{eq1}) is known and linear will be discussed in Section \ref{Sec_lin}.

\section{Quantized Consensus of Uncertain Nonlinear Multi-Agent Systems}\label{Sec_non}

In this section, we present the ESO-based dynamic encoding-decoding scheme with zooming-in and holding capabilities. The control protocol is based on the outputs of the observers, encoders, and decoders. Conditions for quantized consensus under data-rate constraints and DoS attacks will be derived.

\subsection{Protocol Design}

To present the main results, we first conduct a state transformation to the agent dynamics. Let $\rho_i=[\rho_{i1}, \ldots, \rho_{ir}]^{\rm{T}}\in\mathbb{R}^r$ with $\rho_{i1}=y_i$ and $\rho_{ij}=\dot{\rho}_{i,j-1}$, $2\leq j\leq r$. Then by an iterative procedure \cite{Meng-2014}, the agent dynamics (\ref{eq1}) can be transformed into the following normal form \cite{Isi-1989,Khalil-2002}:
\begin{equation}\label{eq32}
   \left\{
  \begin{aligned}
          \dot{\rho}_i=& A_{\rho}\rho_i+B_{\rho}[F_i(\rho_i, z_i, \omega_i)+u_i], \\
                    \dot{z}_i  =& F_{i0}(\rho_i,z_i,\omega_i), \\
                y_i=& \rho_{i1}, ~i=1, \ldots, N,
        \end{aligned} \right.
\end{equation}
where $F_i\in C^1(\mathbb{R}^{r}\times \mathbb{R}^{n_i-r}\times \mathbb{R}^{n_{\omega_i}}, \mathbb{R})$,  $F_{i0}\in C^1(\mathbb{R}^{r}\times \mathbb{R}^{n_i-r}\times \mathbb{R}^{n_{\omega_i}}, \mathbb{R}^{n_i-r})$, and matrices $A_{\rho}\in \mathbb{R}^{r\times r}$ and $B_{\rho}\in\mathbb{R}^{r\times 1}$ are given by
\begin{equation*}
   A_{\rho}=
\left[
  \begin{array}{cccc}
    0 & 1 &  \cdots & 0 \\
    \vdots & \vdots & \ddots & \vdots \\
    0 & 0 &  \cdots & 1 \\
    0 & 0 &    \cdots  &0 \\
  \end{array}
\right], ~B_{\rho}=\left[
             \begin{array}{c}
               0 \\
               \vdots \\
               0 \\
               1 \\
             \end{array}
           \right].
 \end{equation*}

\emph{Lemma 1: For $1\leq i\leq N$, let
\begin{equation}\label{eq2}
h_i=k_1\rho_{i1}+k_{2}\rho_{i2}+\cdots+k_{r-1}\rho_{i,r-1}+\rho_{ir},
\end{equation}
where $k_{1}$, $\ldots$, $k_{r-1}$ are selected such that the polynomial $k_1+k_{2}\lambda+\cdots+k_{r-1}\lambda^{r-1}+\lambda^r$ is Hurwitz. If $h_i$, $1\leq i\leq N$, are bounded and $\lim_{t\rightarrow \infty}h_i=h^*+O(\nu)$ for some constant $h^*$ and small positive constant $\nu$, then $\lim_{t\rightarrow \infty} (y_i-y_j)=O(\nu)$, $1\leq i\neq j\leq N$.}

\emph{Proof:} Let $\tilde{\rho}_{i}=[\rho_{i1},\ldots, \rho_{i,r-1}]^{\rm{T}}\in\mathbb{R}^{r-1}$. From (\ref{eq2}), one has
\begin{equation}\label{eq3}
\dot{\tilde{\rho}}_{i}=\tilde{A}\tilde{\rho}_{i}+\tilde{B}h_i,
\end{equation}
where the matrices $\tilde{A}\in \mathbb{R}^{(r-1)\times (r-1)}$ and $\tilde{B}\in\mathbb{R}^{(r-1)\times 1}$ are given by
\begin{align}\label{eq132}
  \tilde{A}=\left[
            \begin{array}{cccc}
              0 & 1 & \cdots &  0\\
              \vdots & \vdots & \ddots & \vdots \\
              0 & 0 & \cdots &  1\\
              -k_1 & -k_2 & \cdots & -k_{r-1} \\
            \end{array}
          \right], \tilde{B}=\left[
                          \begin{array}{c}
                            0 \\
                            \vdots \\
                            0 \\
                            1 \\
                          \end{array}
                        \right].
\end{align}
Note that system (\ref{eq3}) is a stable linear time-invariant system in controllable canonical form with state $\tilde{\rho}_{i}$ and input $h_i$. Then by Lemma 1 in \cite{Dong-2019} and the superposition principle in linear control theory, one can readily conclude that $\lim_{t\rightarrow \infty} (y_i-y_j)=O(\nu)$, $1\leq i\neq j\leq N$. \IEEEQED

According to Lemma 1, the consensus of the variables $h_i$, $1\leq i\leq N$, implies the output consensus of the multi-agent system (\ref{eq1}). In the sequel, the control protocol $u_i$ will be designed to make $h_i$ achieve consensus. Hence, an attempting in transmitting the information of $h_i$ will be made between each paired agents.  In \cite{You-2011,Feng-2020b}, for higher-order multi-agent systems, the agent state in vector form was transmitted through the digital network.  Recall that $h_i$ is a linear combination of the agent states  and the agents are single-output.  We will show that transmitting the 1-dimensional signal $h_i$ rather than the $r$-dimensional state enables us to achieve the lowest data rate.

\begin{figure*}
   \centering
   \includegraphics[width=0.9\textwidth,bb=18 15 790 355, clip]{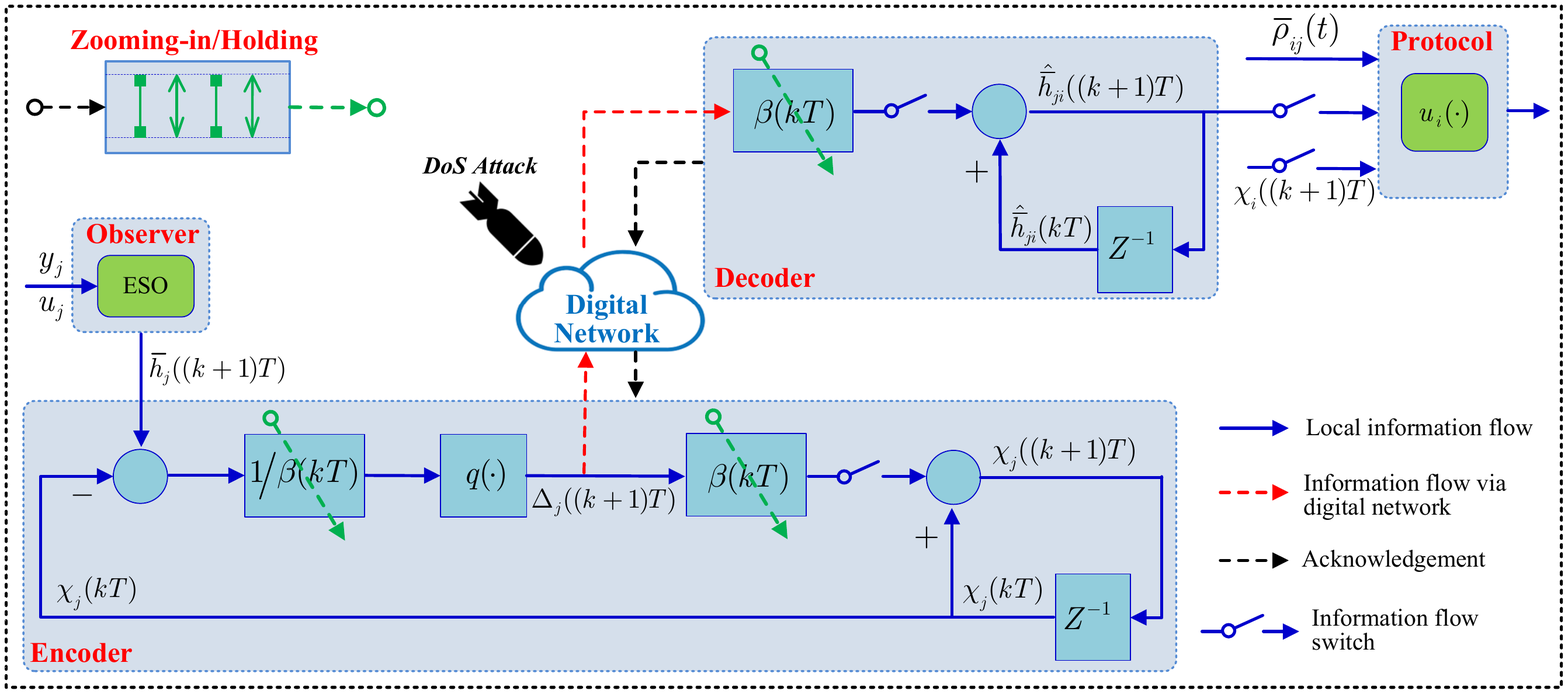}
   \caption{Architecture of the proposed control scheme for uncertain nonlinear multi-agent systems under data-rate constraints and DoS attacks.}\label{fig1}
\end{figure*}

Since for each agent, only the first state variable $y_i=\rho_{i1}$ is available,  the unmeasurable agent states and unknown agent dynamics will be estimated online. Augmenting an extended integrator to the chain of integrators in (\ref{eq32}), the ESO for the extended system is designed as
\begin{equation}\label{eq10}
\left\{
\begin{aligned}
\dot{\hat{\rho}}_i=& A_{\rho}\hat{\rho}_i+B_{\rho}(\hat{\rho}_{i,r+1}+u_i)+L(\varepsilon)(y_i-\hat{\rho}_{i1}), \\
\dot{\hat{\rho}}_{i,r+1}=&\frac{l_{r+1}}{\varepsilon^{r+1}}(y_i-\hat{\rho}_{i1}),
\end{aligned}
\right.
\end{equation}
where $\hat{\rho}_i=[\hat{\rho}_{i1}, \ldots, \hat{\rho}_{ir}]^{\rm{T}}\in\mathbb{R}^{r}$ is the estimate of the agent state $\rho_i$, and $\hat{\rho}_{i,r+1}$ is the estimate of the extended state $\rho_{i,r+1}\triangleq F_i(\rho_i,z_i,\omega_i)$, which represents the total effect of the unknown nonlinear dynamics and external disturbance. For the observer (\ref{eq10}), the initial condition is set as $\textbf{0}_{r+1}$, $\varepsilon<1$ is a small positive constant, $L(\varepsilon)=\left[\frac{l_{1}}{\varepsilon}, \ldots, \frac{l_{r}}{\varepsilon^{r}}\right]^{\rm{T}}$, and  $l_{1}$, $\ldots$, $l_{r+1}$ are selected such that the following matrix is Hurwitz:
\begin{equation}\label{eq34}
  E=\left[
        \begin{array}{ccccc}
          -l_{1} & 1 &  \cdots & 0 \\
          \vdots & \vdots  & \ddots & \vdots  \\
          -l_{r}   & 0  & \cdots  & 1 \\
          -l_{r+1} & 0 & \cdots & 0 \\
        \end{array}
      \right].
\end{equation}

In this paper, we assume that the output $y_i$ is free of measurement noise. Otherwise, some other techniques, such as nonlinear-gain structure \cite{Khalil-2012}, should be adopted to improve the robustness to measurement noise. What is more, the peaking phenomenon is also known as an important feature of the ESO  (\ref{eq10}). More specifically, $ |\rho_{ij}-\hat{\rho}_{ij}|$ , $2\leq j\leq r+1$, might peak to $O(1/\varepsilon^{j-1})$ value during the initial short transient period. To overcome the peaking phenomenon, the saturation technique \cite{Khalil-2017} is employed to saturate the estimates $\hat{\rho}_{i1}$ to $\hat{\rho}_{i,r+1}$ outside a compact set of interest. Let $\overline{\rho}_{ij}=M_{j}\textrm{sat}\left(\frac{\hat{\rho}_{ij}}{M_{j}}\right), ~1\leq j\leq r+1$, where $M_{j}$ are bounds selected such that the saturations will not be invoked during the steady period of the observer  \cite{Khalil-2017}. Correspondingly, denote $\overline{\rho}_i=[\overline{\rho}_{i1}, \ldots, \overline{\rho}_{ir}]^{\rm{T}}$ and
\begin{equation}\label{eq36}
\overline{h}_i=k_1\overline{\rho}_{i1}+k_2\overline{\rho}_{i2}+\cdots+k_{r-1}\overline{\rho}_{i, r-1}+\overline{\rho}_{ir}.
\end{equation}

Based on the output of the ESO, the information of $\overline{h}_i$ will be transmitted through the DoS affected digital network. Recall that $t=kT$ with $k=0, 1, \ldots$ denotes the periodic sampling instants at which the multi-agent system (\ref{eq1}) attempts to exchange information. The data to be transmitted at time $t=kT$ is denoted by $\overline{h}_i(kT)$, and some of them may not be successfully transmitted if DoS is present at $kT$. For the instants at which DoS is not present, as mentioned before Assumption A4, we let $v_0T, v_1T, v_2T, \ldots$ represent the sequence of time instants when the transmissions are successful. Let $H_v\triangleq \{v_i; i\geq 0\}$. According to the acknowledgement signal, the encoder for agent $i$ is designed as
\begin{equation}\label{eq37}
 \left\{
  \begin{aligned}
   \chi_i(0) = & 0, \\
  \chi_i((k+1)T)= & \chi_i(kT)+\beta(kT)\Delta_i((k+1)T), \\
   & \textrm{if}~ k+1 \in H_v,\\
  \chi_i((k+1)T)= & \chi_i(kT),  ~\textrm{if}~ k+1 \notin H_v,\\
  \Delta_i((k+1)T)= & q\left(\frac{\overline{h}_i((k+1)T)-\chi_i(kT)}{\beta(kT)}\right),
  \end{aligned} \right.
\end{equation}
where $\chi_i$ is the internal state of the encoder and $\Delta_i$ is a binary data which represents the output of the encoder. In (\ref{eq37}),  $\beta(kT)$ is the scaling function which is initialized with $\beta(0)=\beta_0>0$ and updated by the following mechanism:
\begin{equation}\label{eq40}
  \beta((k+1)T)=\max\{\gamma \beta(kT), \sqrt{\varepsilon}\},
\end{equation}
with
\begin{equation}\label{eq82}
 \gamma=\left\{
  \begin{aligned}
  \gamma_1,  ~\textrm{if}~ k+1 \in H_v,  \\
  1,~ ~\textrm{if}~ k+1 \notin H_v,
  \end{aligned} \right.
\end{equation}
where $0<\gamma_1<1$ will be specified latter. Note that here we let $\beta(kT)\rightarrow \sqrt{\varepsilon}$ rather than 0 as in \cite{Feng-2020b}. The explanation is that in the quantizer employed in (\ref{eq37}), we use the ESO estimated information $\overline{h}_i((k+1)T)$ rather than the accurate information $h_i((k+1)T)$. In the subsequent section,  we will show that  for any time interval $[\tau_1, \tau_2]\subseteq [0,\infty)$, $\frac{\int_{\tau_1}^{\tau_2}\left(h_i(\tau)-\overline{h}_{i}(\tau)\right)\textrm{d}\tau}{\sqrt{\varepsilon}}\to 0$ as  $\varepsilon\to 0$. Therefore, letting $\beta(kT)\rightarrow \sqrt{\varepsilon}$ prevents the saturation of the quantizer from the accumulated ESO estimation error. What is more, note that in (\ref{eq37}), the scaling function $\beta(kT)$ appears in the denominator of the information to be quantized. This together with its update mechanism  (\ref{eq40})-(\ref{eq82}), for $\beta(kT)> \sqrt{\varepsilon}$, the sensitivity of the quantizer to the term  $\left[\overline{h}_i((k+1)T)-\chi_i(kT)\right]$ increases if the system is free of  DoS attacks (since $\gamma=\gamma_1<1$), and keeps invariant if under DoS attacks (since $\gamma=1$).

If agent $i$ is a neighbor of agent $j$, $\Delta_j$ (which is the output of the encoder for agent $j$) will be transmitted through the DoS affected digital network. Agent $i$ uses the following decoder to reconstruct $\overline{h}_j$:
\begin{equation}\label{eq38}
 \left\{
  \begin{aligned}
  \hat{\overline{h}}_{ji}(0)=&  0, \\
  \hat{\overline{h}}_{ji}((k+1)T)= & \hat{\overline{h}}_{ji}(kT)+\beta(kT)\Delta_j((k+1)T), \\
  & \textrm{if}~ k+1 \in H_v,\\
  \hat{\overline{h}}_{ji}((k+1)T) = & \hat{\overline{h}}_{ji}(kT), ~\textrm{if}~ k+1 \notin H_v,
  \end{aligned} \right.
\end{equation}
where $\hat{\overline{h}}_{ji}$ is the estimate of $\overline{h}_j$ obtained by agent $i$. Since the initial conditions of the encoders and decoders are set as 0, we let $v_0=0$ henceforth.

With the aid of the ESO (\ref{eq10}), encoder (\ref{eq37}), and decoder (\ref{eq38}), we propose the following protocol:
\begin{align}\label{eq39}
 u_i(t)= & -\sum_{\ell=1}^{r-1}k_{\ell}\overline{\rho}_{i,\ell+1}(t)-\overline{\rho}_{i,r+1}(t) \nonumber \\
 & +c\sum_{j\in\mathcal{N}^{\textrm{DoS}}_i(kT)}a_{ij}\left(\hat{\overline{h}}_{ji}(kT)-\chi_i(kT)\right), \nonumber \\
 &  t\in [kT, (k+1)T),
 \end{align}
where $c>0$, and $\mathcal{N}_i^{\textrm{DoS}}(kT)$ represents the set of agents which are able to exchange information with agent $i$ through the DoS affected digital network at $t=kT$. That is, $\mathcal{N}_i^{\textrm{DoS}}(kT)=\mathcal{N}_i$ if $k\in H_v$, and $\mathcal{N}_i^{\textrm{DoS}}(kT)=\emptyset$ otherwise. In the control protocol (\ref{eq39}), the term $-\sum_{\ell=1}^{r-1}k_{\ell}\overline{\rho}_{i,\ell+1}(t)-\overline{\rho}_{i,r+1}(t)$ is to compensate for the agent dynamics in real time, and the term $c\sum_{j\in\mathcal{N}^{\textrm{DoS}}_i(kT)}a_{ij}\left(\hat{\overline{h}}_{ji}(kT)-\chi_i(kT)\right)$  satisfies
\begin{align*}
 & c\sum_{j\in\mathcal{N}^{\textrm{DoS}}_i(kT)}a_{ij}\left(\hat{\overline{h}}_{ji}(kT)-\chi_i(kT)\right)  \\
 =&  c\sum_{j\in\mathcal{N}^{\textrm{DoS}}_i(kT)}a_{ij}\left(\overline{h}_j(kT)-\overline{h}_i(kT)\right)  \\
   &  -c\sum_{j\in\mathcal{N}^{\textrm{DoS}}_i(kT)}a_{ij}\left(\overline{h}_j(kT)-\hat{\overline{h}}_{ji}(kT)\right) \\
   & +c\sum_{j\in\mathcal{N}^{\textrm{DoS}}_i(kT)}a_{ji}\left(\overline{h}_i(kT)-\hat{\overline{h}}_{ij}(kT)\right),
\end{align*}
where the second term represents the weighted sum of the estimation errors for agent $i$'s neighbors' information $\overline{h}_j(kT)$; and the last term represents the weighted sum of the estimation errors for $\overline{h}_i(kT)$ by agent $i$’s neighbors. The last term plays an error-compensation role in the protocol and is mainly based on the  assumption that $a_{ij}=a_{ji}$. This mechanism was first considered in \cite{Boyd-2004}, and also employed in many previous quantized consensus results \cite{Carli-2010,Li-2011,Qiu-2016,You-2011,Meng-2017,Dong-2019}. The extension of the protocol (\ref{eq39}) to direct communication graphs is not straightforward, and some extra techniques are needed \cite{Lin-2013}.

The proposed control scheme for uncertain nonlinear multi-agent systems under data-rate constraints and DoS attacks is depicted in Fig. \ref{fig1}.

\subsection{Convergence Analysis}\label{Sec_3b}

Before stating the convergence results, we first present the dynamics of the multi-agent system in the absence and presence of DoS attacks. Denote $H(kT)  =[h_1(kT),  \ldots, h_N(kT)]^{\rm{T}}$, $\overline{H} (kT)  =[\overline{h}_1(kT),  \ldots, \overline{h}_N(kT)]^{\rm{T}}$, $\widehat{\overline{H}}(kT) =[\chi_1(kT),\ldots, \chi_N(kT)]^{\rm{T}}$, $\mu(kT) = \overline{H} (kT)-\widehat{\overline{H}}(kT)$, and $\delta(kT)  = H(kT)-J_NH(kT)$.
Note that $\mu(kT)$ and $\delta(kT)$ represent the compact quantization error and  consensus error, respectively.

If the multi-agent system is \emph{in the absence of DoS attack}, i.e., $k\in H_v$, then by (\ref{eq39}) and the definition of $h_i$, one has
\begin{align}\label{eq12}
 \dot{h}_i(t) = & \sum_{\ell=1}^{r}k_{\ell}\rho_{i, \ell+1}(t)+u_i(t) \nonumber \\
 =& c\sum_{j\in\mathcal{N}_i}a_{ij}\left(\hat{\overline{h}}_{ji}(kT)-\chi_i(kT)\right) \nonumber \\
 & +\sum_{\ell=1}^{r}k_{\ell}\left(\rho_{i, \ell+1}(t)-\overline{\rho}_{i,\ell+1}(t)\right), t\in[kT, (k+1)T), \nonumber \\
 &
\end{align}
where $k_{r}=1$. It follows that
\begin{align}\label{eq13}
 & h_i((k+1)T)=  h_i(kT) \nonumber \\
 & \qquad + cT\sum_{j\in\mathcal{N}_i}a_{ij}\left(\hat{\overline{h}}_{ji}(kT)-\chi_i(kT)\right)+\varsigma_{1i}(kT),
\end{align}
where
\begin{align*}
\varsigma_{1i}(kT)= \int_{kT}^{(k+1)T}\left(\sum_{\ell=1}^{r}k_{\ell}\left(\rho_{i,\ell+1}(t)-\overline{\rho}_{i,\ell+1}(t)\right)\right)\textrm{d}t.
\end{align*}
According to the structures of the paired encoder (\ref{eq37}) and decoder (\ref{eq38}), one has
\begin{equation}\label{eq14}
  \hat{\overline{h}}_{ji}(kT)=\chi_j(kT), ~i\in\mathcal{N}_j, ~j=1, 2, \ldots, N.
\end{equation}
This together with (\ref{eq13}) leads to
\begin{align}\label{eq15}
H((k+1)T) = & H(kT)-cT\mathcal{L}\widehat{\overline{H}}(kT)+\varsigma_1(kT) \nonumber \\
= & (I-cT\mathcal{L})\overline{H}(kT)+cT\mathcal{L}\mu(kT)\nonumber \\
    & +H(kT)-\overline{H}(kT) +\varsigma_1(kT),
\end{align}
where $\varsigma_1(kT)=[\varsigma_{11}(kT), \ldots, \varsigma_{1N}(kT)]^{\textrm{T}}$. Since $\mathcal{L}J_N=J_N\mathcal{L}=0$, by (\ref{eq15}), one gets
\begin{align}\label{eq17}
\delta((k+1)T) = & (I-J_N)H((k+1)T) \nonumber \\
= & (I-cT\mathcal{L})\delta(kT)+cT\mathcal{L}\mu(kT)  \nonumber \\
  & +cT\mathcal{L}\left(H(kT)-\overline{H}(kT)\right)\nonumber \\
  & +(I-J_N)\varsigma_1(kT),
\end{align}
and
\begin{align}\label{eq71}
  &  \overline{H}((k+1)T)-\widehat{\overline{H}}(kT)  \nonumber \\
=   &  H((k+1)T)-\widehat{\overline{H}}(kT)+\overline{H}((k+1)T)-H((k+1)T) \nonumber \\
 =& (I+cT\mathcal{L})\mu(kT)-cT\mathcal{L}\delta(kT) \nonumber \\
   & +(I+cT\mathcal{L})(H(kT) -\overline{H}(kT)) \nonumber \\
 &+\varsigma_1(kT)+\overline{H}((k+1)T)-H((k+1)T).
\end{align}

If the multi-agent system is \emph{in the presence of DoS attack}, i.e., $k\notin H_v$, then one has
\begin{equation}\label{eq5}
 \dot{h}_i(t) = \sum_{\ell=1}^{r}k_{\ell}\left(\rho_{i, \ell+1}(t)-\overline{\rho}_{i,\ell+1}(t)\right), t\in[kT, (k+1)T).
\end{equation}
It follows that
\begin{align}\label{eq56}
H((k+1)T) =  H(kT)+\varsigma_{1}(kT).
\end{align}
By (\ref{eq56}), one has
\begin{align}\label{eq79}
    \delta((k+1)T)=\delta(kT)+(I-J_N)\varsigma_{1}(kT),
\end{align}
and
\begin{align}\label{eq57}
  &  \overline{H}((k+1)T)-\widehat{\overline{H}}(kT)  \nonumber \\
 =& \mu(kT)+\varsigma_1(kT)+H(kT)-\overline{H}(kT)\nonumber \\
 & +\overline{H}((k+1)T)-H((k+1)T).
\end{align}

According to the update mechanisms of the encoders and decoders, the evolution equations of $\mu(kT)$ are stated as follows:
\begin{align}
 \mu((k+1)T)=  & \left[\overline{H}((k+1)T)-\widehat{\overline{H}}(kT)\right] \nonumber \\
 & -\beta(kT) Q\left(\frac{\overline{H}((k+1)T)-\widehat{\overline{H}}(kT)}{\beta(kT)}\right), \nonumber \\
\label{eq54} & \textrm{if} ~ k+1\in H_v, \\
 \mu((k+1)T)=  & \overline{H}((k+1)T)-\widehat{\overline{H}}(kT),  \nonumber \\
\label{eq88} & \textrm{if} ~ k+1\notin H_v,
\end{align}
where $\left[\overline{H}((k+1)T)-\widehat{\overline{H}}(kT)\right]$ is given by (\ref{eq71}) if $k\in H_v$, and given by (\ref{eq57}) if $k\notin H_v$.

The dynamics of the multi-agent system, in terms of the quantization error $\mu(kT)$ and consensus error $\delta(kT)$, are given by (\ref{eq17}), (\ref{eq79}), (\ref{eq54}), and (\ref{eq88}). The following remark explains the design philosophy of the proposed zooming-in and holding approach in the context of quantized consensus of multi-agent systems under DoS attacks.

\emph{Remark 2 (Design Philosophy):} When the transmissions are successful, the agents are driven to achieve consensus.  In this case, similar to \cite{Li-2011,Lin-2013,Qiu-2016,You-2011,Meng-2017,Dong-2019,Feng-2020b}, the scaling function $\beta(kT)$ zooms-in the information to be quantized (i.e., the term $\left[\overline{h}_i((k+1)T)-\chi_i(kT)\right]$) to decrease the consensus error. Therefore, $\gamma_1$ is selected less than 1. When the transmissions are unsuccessful, the consensus process is interrupted. The strategy in this paper is to hold the previous status of the whole system. In the consensus error dynamics (\ref{eq79}) under DoS attacks, anticipating that $\varsigma_1(kT)\rightarrow 0$ as $\varepsilon\rightarrow 0$, one gets $\delta((k+1)T)=\delta(kT)$ if $k\notin H_v$. What is more, the status of the encoders and decoders are also hold, as we let $\chi_i(kT)=\chi_i((k-1)T)$, $\hat{\overline{h}}_{ji}(kT) =  \hat{\overline{h}}_{ji}((k-1)T)$, and $\beta(kT)=\beta((k-1)T)$ if $k\notin H_v$. Such a zooming-in and holding mechanism is expected to lead to higher tolerance to DoS attacks. We should point out that the zooming-in and holding approach is fundamentally different from the  zooming-in and zooming-out approach in \cite{Feng-2020b} (see Section \ref{Sec_lin} for more details). \IEEEQED

In the sequel, we will state the convergence results of the closed-loop system. Let $\mathcal{X}^{\dag}$ be a compact subset of $\mathbb{R}^{r}$, and $\mathcal{X}$  slightly smaller than $\mathcal{X}^{\dag}$ (i.e., $\mathcal{X}\subseteq \mathcal{X}^{\dag}$ and their boundaries are disjoint). Denote $C_h\geq \max_{1\leq i\leq N}|h_i(0)|$. Let $\omega_i\in \mathcal{W}_i$ for some compact set $\mathcal{W}_i\subseteq\mathbb{R}^{n_{\omega_i}}$, and let $z_i(0)\in \mathcal{Z}_i$ for some compact set $\mathcal{Z}_i\subseteq\mathbb{R}^{n_i-r}$. By Assumption A3, there exists a positive constant $N_{zi}$ such that $\sup_{t\in[0,\infty)}\|z_i(t)\|\leq N_{zi}$ for all $z_i(0)\in \mathcal{Z}_i$, $\rho_i\in \mathcal{X}^{\dag}$, and $\omega_i\in\mathcal{W}_i$. Note that any compact subset of $\mathbb{R}^{r}\times \mathbb{R}^{n_i-r}$ can be put in the interior of $\mathcal{X}\times \mathcal{Z}_i$. We first state the following lemma.

\emph{Lemma 2 \cite{Li-2011}: If Assumption A1 holds and $c\in(0, 2/(T\lambda_N))$, then $\rho_h<1$, where
\begin{equation}\label{eq51}
  \rho_h=\max_{2\leq i\leq N}\left|1-cT\lambda_i\right|.
\end{equation}
Furthermore, if  $c\in (0, 2/(T(\lambda_2+\lambda_N)))$, then $\rho_h=1-cT\lambda_2$.}

Now, we are in a position to state our first main result.

\emph{\textbf{Theorem 1:} Consider the multi-agent system (\ref{eq1}) with a digital communication network subject to data-rate constraints and DoS attacks. Given the control protocol (\ref{eq39}) with ESO (\ref{eq10}), encoder (\ref{eq37}), and decoder (\ref{eq38}). Suppose Assumptions A1 to A4 are satisfied, and the initial conditions of the agents $(\rho_i(0),z_i(0))\in \mathcal{X}\times \mathcal{Z}_i$, $1\leq i\leq N$. Let
\begin{align}
\label{eq23} c &   \in \left(0, 2/(T\lambda_N)\right),  \\
\label{eq24} \gamma_1 &   \in  (\rho_h, 1), \\
\label{eq25}    K & \geq  \left\lfloor K_1(c,\gamma_1)-\frac{1}{2}   \right\rfloor+1, \\
\label{eq28} \beta_0 & > \max\left\{\frac{2cT\lambda_NC_{h}}{\gamma_1(K+\frac{1}{2})}, \frac{4C_h\gamma_1(\gamma_1-\rho_h)}{cT\lambda_N} \right\},
\end{align}
where
\begin{align}
\label{eq27} & K_1(c, \gamma_1)= \frac{1+2cTd^*}{2\gamma_1}+\frac{\sqrt{N}c^2T^2\lambda^2_N}{2\gamma_1(\gamma_1-\rho_h)}.
 \end{align}
Then for any $\sigma>0$, there exists $\varepsilon^{\dag}>0$, which is dependent on $\sigma$, such that $\forall \varepsilon\in(0,\varepsilon^{\dag})$:
\begin{itemize}
  \item the ESO (\ref{eq10}) achieves practical convergence, i.e., there exists $\tau_0(\varepsilon)>0$ satisfying $\lim_{\varepsilon\rightarrow 0}\tau_0(\varepsilon)=0$, such that $\forall t\in[\tau_0(\varepsilon),\infty)$,
     \begin{equation}\label{eq22}
        \left|\rho_{ij}(t)-\hat{\rho}_{ij}(t)\right|\leq \sigma, ~1\leq i\leq N, ~1\leq j\leq r+1;
     \end{equation}
  \item the quantizer will never be saturated, and the multi-agent system (\ref{eq1}) achieves practical output consensus, i.e.,
    \begin{equation}\label{eq41}
        \lim_{t\rightarrow \infty}\left|y_i(t)-y_j(t)\right|\leq \sigma, ~ 1\leq i\neq j\leq N.
    \end{equation}
\end{itemize}}

\emph{Proof:} See the Appendix. \IEEEQED

As stated in Theorem 1, the DoS attacks can be arbitrarily serious as long as their induced maximal number of consecutive packet losses is bounded. Also note that $K\rightarrow \infty$ as $N\rightarrow \infty$. However, the quantization level (or data-rate) of a real digital network is limited. Our next theorem shows that the selection of $K$ can be made independent of the number of the agents. In particular, the lowest quantization level we achieve is 3, i.e., $K=1$. That is, the communication channel is required to be capable of transmitting merely 2 bits at each time step, no matter how large the number of the agents is.

\emph{\textbf{Theorem 2:} Consider the multi-agent system (\ref{eq1}) with a digital communication network subject to data-rate constraints and DoS attacks. Given the control protocol (\ref{eq39}) with ESO (\ref{eq10}), encoder (\ref{eq37}), and decoder (\ref{eq38}). Suppose Assumptions A1 to A4 are satisfied, and the initial conditions of the agents $(\rho_i(0),z_i(0))\in \mathcal{X}\times \mathcal{Z}_i$, $1\leq i\leq N$. For any given $K\geq 1$, let $\beta_0$ be selected according to (\ref{eq28}), and let
\begin{align}
\label{eq97}   c  \in &(0, \min\{2/(T(\lambda_2+\lambda_N)), c_m\}),  \\
\label{eq98}   \gamma_1= & 1-(1-\epsilon_0)cT\lambda_2,
\end{align}
where $\epsilon_0\in(0,1)$ and
$c_m=2K\epsilon_0\lambda_2T^{-1}\left[\sqrt{N}\lambda^2_N+2\epsilon_0\lambda_2d^*  \right.$ $\left. +(2K+1)(1-\epsilon_0)\epsilon_0\lambda_2^2\right]^{-1}$. Then the statements in Theorem 1 hold.}

\emph{Proof:} By the definition of $c_m$, one can verify that $K_1(c,\gamma_1)$ specified by (\ref{eq27}) satisfies $\frac{1}{2}<K_1(c,\gamma_1)<K+\frac{1}{2}$. This together with (\ref{eq97}) and $(\ref{eq98})$ yields that the conditions in (\ref{eq23})-(\ref{eq25}) are all satisfied. Then, similar to Theorem 1, one can complete the proof of Theorem 2. \IEEEQED

Theorem 2 indicates that the proposed control protocol is capable of handling any DoS attacks inducing bounded consecutive packet losses for the uncertain nonlinear multi-agent system (\ref{eq1}) with merely 2 bits information exchange between each pair of adjacent agents for each transmission attempt.  Also note that $\gamma_1\rightarrow 1$ as $N\rightarrow \infty$,  which means that, under limited data rate, the convergence rate becomes slow if the number of the agents increases.

\emph{Remark 3 (Parameters Selection):} The saturation bounds $M_1$ to $M_{r+1}$ satisfying $M_j\geq \sup_{1\leq i\leq N}|\rho_{ij}|$, $1\leq j\leq r+1$, are selected such that the saturations will not be invoked in the steady state of the observer \cite{Khalil-2017}.  The parameter $C_h$ is to  prevent the saturation of the quantizer in the initial stage \cite{Li-2011,Lin-2013,Qiu-2016,You-2011,Meng-2017,Dong-2019,Feng-2020b}. The selections of $M_j$ and $C_h$ might end up with some conservative values due to the uncertain nonlinear agent dynamics. For the parameter $\varepsilon$, theoretically, smaller $\varepsilon$ leads to higher estimation and consensus accuracy. Unfortunately, in case of uncertain dynamics, concrete and constructive algorithms to calculate the upper bound of $\varepsilon$ (i.e., $\varepsilon^{\dag}$) are very challenging to obtain and still missing in the literature \cite{Khalil-2017}. This is indeed a cost we pay for the uncertainties. In this paper, the scope of the current work is to develop a control scheme with the existence of certain parameters to handle the quantized consensus problem under data-rate constraints and DoS attacks. In practice, such parameters can be decided by a simple trial and error procedure \cite{Khalil-2017}, or by a simple line search algorithm \cite{Valm-2010}. In fact, the ESO has been successfully adopted in many control engineering problems in recent years  \cite{Chen-2016,Sari-2020}. At last, we mention that the sampling period $T$ is not a design parameter in Theorem 1. As $T$ is always multiplied by the constant $c$, the effects of the sampling period on the control design can be adjusted  by the value of $c$.
\IEEEQED

\emph{Remark 4 (Comparison with \cite{Ran-2019}): } In our previous work \cite{Ran-2019}, a higher-order ESO-based protocol was developed to handle the limited data rate problem for a class of simpler nonlinear multi-agent systems without DoS attacks. In this paper, we consider more general uncertain nonlinear systems, and our focus is  on the coexistence problem of data-rate constraints and DoS attacks. Due to the communication interruption caused by DoS,  the consensus error may diverge during the DoS periods, while in \cite{Ran-2019} the communication network is always reliable. Therefore, in this paper new encoding-decoding scheme is needed, and the behavior of the system during  the attack periods must be carefully governed.  Accordingly, we developed a novel dynamic quantization with zooming-in and holding capabilities, and showed that it enables better transient consensus performance and higher level of tolerable DoS attacks. These render the contribution of this paper with respect to \cite{Ran-2019} substantial.
\IEEEQED

\section{Quantized Consensus of Known Linear Multi-Agent Systems}\label{Sec_lin}

In this section, for the ease of comparison with state-of-the-art results, we revisit the quantized consensus problem for a class of known discrete-time linear multi-agent systems under data-rate constraints and DoS attacks.

\subsection{Protocol Design}

Without loss of generality, it is assumed that the model of agent $i$, $1\leq i\leq N$, is already in the following controllable canonical form:
\begin{equation}\label{eq99}
   x_i((k+1)T)=Ax_i(kT)+Bu_i(kT),
\end{equation}
where $x_i\in \mathbb{R}^r$ is the state, and matrices $A\in\mathbb{R}^{r\times r}$ and $B\in\mathbb{R}^{r\times 1}$ are given by
\begin{align*}
  A=\left[
            \begin{array}{cccc}
              0 & 1 & \cdots &  0\\
              \vdots & \vdots & \ddots & \vdots \\
              0 & 0 & \cdots &  1\\
              -a_1 & -a_{2} & \cdots & -a_{r} \\
            \end{array}
          \right], B=\left[
                          \begin{array}{c}
                            0 \\
                            \vdots \\
                            0 \\
                            1 \\
                          \end{array}
                        \right].
\end{align*}
Here we are interested in the case where $A$ has at least one eigenvalue on or outside the unit circle. Otherwise, the multi-agent system (\ref{eq99}) can achieve state consensus by setting $u_i(kT)\equiv 0$. This implies that any communication protocol among the agents and any pattern of DoS attacks do not influence the trivial consensus problem if $A$ is a stable matrix.

Denote
\begin{equation}\label{eq16}
h_i(kT)=k_1x_{i1}(kT)+\cdots+k_{r-1}x_{i, r-1}(kT)+x_{ir}(kT).
\end{equation}
For the discrete-time case, $k_{1}$, $\ldots$, $k_{r-1}$ are selected such that the eigenvalues of matrix $\tilde{A}$ specified by (\ref{eq132}) are real and distinct, and located inside the unit circle. For subsequent use, let $\tilde{\lambda}_1$, $\ldots$, $\tilde{\lambda}_{r-1}$ be the eigenvalues of matrix $\tilde{A}$, and let $\tilde{\Lambda}=\textrm{diag}\{\tilde{\lambda}_1, \ldots, \tilde{\lambda}_{r-1}\}$. It follows that there exists an invertible matrix $\Gamma\in \mathbb{R}^{(r-1)\times (r-1)}$ such that $\Gamma\tilde{A}\Gamma^{-1}=\tilde{\Lambda}$. Let $\tilde{\Lambda}^{\dag}=\textrm{diag}\{1/(1-\tilde{\lambda}_1), \ldots, 1/(1-\tilde{\lambda}_{r-1})\}$ and $\tilde{A}_{r-1}=[-k_1, \ldots, -k_{r-1}]$ be the $(r-1)$th row of matrix $\tilde{A}$.

\emph{Lemma 3: If $h_i(kT)$, $1\leq i\leq N$, $k\geq 0$, are bounded and $\lim_{k\rightarrow \infty}h_i(kT)=h^*$ for some constant $h^*$, then
\begin{equation}\label{eq102}
  \lim_{k\rightarrow \infty}\|x_i(kT)-x_j(kT)\|= 0, ~1\leq i\neq j\leq N.
\end{equation}
Furthermore,
\begin{equation}\label{eq129}
 \lim_{k\rightarrow \infty}x_i(kT)=
\left[
\begin{aligned}
& ~~~~  \Gamma \tilde{\Lambda}^{\dag} \Gamma^{-1}\tilde{B} \\
& 1+\tilde{A}_{r-1}\Gamma \tilde{\Lambda}^{\dag} \Gamma^{-1}\tilde{B}
 \end{aligned}
\right]h^*,
\end{equation}
where the matrices $\tilde{A}$ and $\tilde{B}$ are given by (\ref{eq132}).}

\emph{Proof:} Let $\tilde{x}_i(kT)=[x_{i1}(kT),\ldots, x_{i,r-1}(kT)]^{\rm{T}}\in\mathbb{R}^{r-1}$. By (\ref{eq16}), one has
\begin{equation}\label{eq128}
\tilde{x}_i((k+1)T)=\tilde{A}\tilde{x}_i(kT)+\tilde{B}h_i(kT).
\end{equation}
Note that the eigenvalues of matrix $\tilde{A}$ are locate inside the unit circle, and $h_i(kT)$ is bounded and converges to $h^*$. From the linear control theory \cite{Chen-1998}, one can readily obtain (\ref{eq102}). Furthermore, according to the response of the discrete-time linear system (\ref{eq128}) and the relation $\tilde{A}^k=\Gamma \tilde{\Lambda}^k \Gamma^{-1}$, one has $\lim_{k\rightarrow \infty}\tilde{x}_i(kT)=\Gamma \tilde{\Lambda}^{\dag} \Gamma^{-1}\tilde{B}h^*$. This together with (\ref{eq16}) yields (\ref{eq129}). \IEEEQED

By Lemma 3,  the consensus of $h_i(kT)$ implies the consensus of $x_i(kT)$. The dynamics of $h_i(kT)$ is given by
\begin{align}\label{eq103}
  h_i((k+1)T)= & \sum_{\ell=1}^{r-1}k_{\ell}x_{i,\ell+1}(kT)-\sum_{m=1}^r a_mx_{im}(kT) \nonumber \\
  & +u_i(kT).
\end{align}

Similar to Section \ref{Sec_non}, the encoder for agent $i$ is designed as
\begin{equation}\label{linear encoder}
 \left\{
  \begin{aligned}
   \chi_i(0) = & 0, \\
  \chi_i((k+1)T)= & \chi_i(kT)+\beta(kT)\Delta_i((k+1)T), \\
   & \textrm{if}~ k+1 \in H_v,\\
  \chi_i((k+1)T)= & \chi_i(kT),  ~\textrm{if}~ k+1 \notin H_v,\\
  \Delta_i((k+1)T)= & q\left(\frac{h_i((k+1)T)-\chi_i(kT)}{\beta(kT)}\right),
  \end{aligned} \right.
\end{equation}
where the scaling function is updated by
\begin{equation}\label{eq100}
  \beta((k+1)T)=\gamma \beta(kT),
\end{equation}
with $\gamma$ given by (\ref{eq82}).  Note that in (\ref{linear encoder}) we use the accurate value of $h_i$, while the ESO estimated value $\overline{h}_i$ in Section \ref{Sec_non}. That is, no additional ESO estimation error will be injected into  the quantization process here. Therefore, in this section we let the scaling function converge to 0 rather than $\sqrt{\varepsilon}$ . The decoder for agent $i$ corresponding to the edge $(i,j)$ is as follows:
\begin{equation}\label{linear_decoder}
 \left\{
  \begin{aligned}
  \hat h_{ji} (0)=&  0, \\
  \hat h_{ji}((k+1)T)= & \hat h_{ji}(kT)+\beta(kT)\Delta_j((k+1)T), \\
  & \textrm{if}~ k+1 \in H_v,\\
  \hat  h_{ji}((k+1)T) = & \hat  h_{ji}(kT), ~\textrm{if}~ k+1 \notin H_v,
  \end{aligned} \right.
\end{equation}
where $\hat{h}_{ji}$ is the estimate of $h_j$  obtained by agent $i$.  Here we remind that for the known linear multi-agent systems in this section, $h_j$ is specified by (\ref{eq16}). The protocol for the known linear multi-agent system (\ref{eq99}) is then given by
\begin{align}\label{eq101}
 u_i(kT)= & -\sum_{\ell=1}^{r-1}k_{\ell}x_{i,\ell+1}(kT)+\sum_{m=1}^r a_mx_{ij}(kT)+h_i(kT) \nonumber \\
 & +cT\sum_{j\in\mathcal{N}^{\textrm{DoS}}_i(kT)}a_{ij}\left(\hat{h}_{ji}(kT)-\chi_i(kT)\right),
 \end{align}
 where $c>0$.

\subsection{Convergence Analysis}

First of all, denote $H(kT)$ and $\delta(kT)$ as in Section \ref{Sec_3b}, and let $\widehat{H}(kT)=[\chi_1(kT),  \ldots, \chi_N(kT)]^{\rm{T}}$ and $\mu(kT)= H(kT)-\widehat{H}(kT)$ in this subsection.

If the multi-agent system (\ref{eq99}) is \emph{in the absence of DoS attack}, i.e., $k\in H_v$, then by (\ref{eq103}) and (\ref{eq101}), one has
\begin{align}\label{eq106}
 h_i((k+1)T)= h_i(kT) +  cT\sum_{j \in \mathcal{N}_i} a_{ij}(\hat h_{ji}(kT)- \chi_i(kT)).
\end{align}
This together with  $\hat{h}_{ji}(kT)=\chi_j(kT)$, $i\in\mathcal{N}_j$, gives
\begin{align}\label{eq107}
H((k+1)T) = & H(kT)-cT\mathcal{L}\widehat{H}(kT) \nonumber \\
= & (I-cT\mathcal{L})H(kT)+cT\mathcal{L}\mu(kT).
\end{align}
By $\mathcal{L}J_N=J_N\mathcal{L}=0$, one has
\begin{align}\label{eq108}
\delta((k+1)T) = (I-cT\mathcal{L})\delta(kT)+cT\mathcal{L}\mu(kT),
\end{align}
and
\begin{align}\label{eq109}
  H((k+1)T)-\widehat{H}(kT)= (I+cT\mathcal{L})\mu(kT)-cT\mathcal{L}\delta(kT).
\end{align}

If the multi-agent system (\ref{eq99}) is \emph{in the presence of DoS attack}, i.e., $k \notin H_v$, then one has
\begin{equation}\label{eq110}
  h_i((k+1)T) =  h_i(kT).
\end{equation}
It follows that
\begin{align}\label{eq112}
    \delta((k+1)T)=\delta(kT),
\end{align}
and
\begin{align}\label{eq113}
  &  H((k+1)T)-\widehat{H}(kT)=\mu(kT).
\end{align}

The evolution equations of $\mu(kT)$ are given by
\begin{align}
 \mu((k+1)T)=  & \left[H((k+1)T)-\widehat{H}(kT)\right] \nonumber \\
 & -\beta(kT) Q\left(\frac{H((k+1)T)-\widehat{H}(kT)}{\beta(kT)}\right), \nonumber \\
\label{eq114} & \textrm{if} ~ k+1\in H_v, \\
\label{eq115} \mu((k+1)T)=  & \mu(kT), ~\textrm{if} ~ k+1\notin H_v,
\end{align}
where $\left[H((k+1)T)-\widehat{H}(kT)\right]$ is given by (\ref{eq109}) if $k\in H_v$, and given by (\ref{eq113}) if $k\notin H_v$.

The dynamics of the multi-agent system (\ref{eq99}), in terms of the quantization error $\mu(kT)$ and consensus error $\delta(kT)$, are represented by (\ref{eq108}), (\ref{eq112}), (\ref{eq114}), and (\ref{eq115}). Now we are ready to present the result concerning the state consensus of the known linear multi-agent system (\ref{eq99}) with the zooming-in and holding approach.

\emph{\textbf{Theorem 3:} Consider the known linear multi-agent system (\ref{eq99}) with a digital network subject to data-rate constraints and DoS attacks. Given the control protocol (\ref{eq101}) with encoder (\ref{linear encoder}) and decoder (\ref{linear_decoder}). Suppose Assumptions A1 and A4 are satisfied, and the initial conditions of the agents $x_i(0)\in \mathcal{X}$, $1\leq i\leq N$. For any given $K\geq 1$, let the parameter $c$ and $\gamma_1$ be selected as in Theorem 2, and the scaling function be updated by (\ref{eq100}) and (\ref{eq82}), and
\begin{align}
\label{eq134} \beta_0 & > \max\left\{\frac{C_{h}}{K+\frac{1}{2}}, \frac{2C_h(\gamma_1-\rho_h)(cT\lambda_N+2\gamma_1)}{cT\lambda_N} \right\}.
\end{align}
Then the quantizer will never be saturated, and the multi-agent system (\ref{eq99}) achieves state consensus specified by (\ref{eq102}). Furthermore, the final consensus value is given by
\begin{equation}\label{eq127}
  \lim_{k\rightarrow \infty}x_i(kT)=
  \left[\begin{aligned}
& ~~~~  \Gamma \tilde{\Lambda}^{\dag} \Gamma^{-1}\tilde{B} \\
& 1+\tilde{A}_{r-1}\Gamma \tilde{\Lambda}^{\dag} \Gamma^{-1}\tilde{B}
 \end{aligned}\right]
 \frac{1}{N}\sum_{m=1}^{N}\sum_{j=1}^{r}k_jx_{mj}(0).
\end{equation}}

\emph{Proof:}  The proof is straightforward by following a similar technique line as in the Proposition 2 in the proof of Theorem 1. Here we omit the details due to space limitation. The initial condition of the scaling function $\beta_0$ is given by (\ref{eq134}) rather than (\ref{eq28}) is because that in this section $\|\mu(0)\|_{\infty}=\|H(0)-\widehat{H}(0)\|_{\infty}\leq C_h$ while in Section \ref{Sec_non} $\|\mu(0)\|_{\infty}=\|\overline{H}(0)-\widehat{\overline{H}}(0)\|_{\infty}=0$. \IEEEQED

\subsection{Discussion}

In this subsection, we first briefly introduce the approach in \cite{Feng-2020b} for comparison purpose. In \cite{Feng-2020b}, a control scheme which adopts a zooming-in and zooming-out mechanism was proposed to tackle the quantized consensus problem of known linear multi-agent systems under data-rate constraints and DoS attacks. The basic idea of this mechanism is to update the encoder as an open-loop system during DoS attacks. In this case, the scaling function needs to zoom-out the information to be quantized to prevent the quantizer being saturated. Specifically, for the multi-agent system (\ref{eq99}), the encoder in \cite{Feng-2020b} is given by
\begin{equation}\label{eq124}
 \left\{
  \begin{aligned}
   \hat{x}_i(0) = & \, \textbf{0}_r, \\
  \hat{x}_i((k+1)T)= &\, A \hat{x}_i(kT)+\beta(kT)\Delta_i((k+1)T), \\
   & \textrm{if}~ k+1 \in H_v,\\
  \hat{x}_i((k+1)T)= &\, A\hat{x}_i(kT),  ~\textrm{if}~ k+1 \notin H_v,\\
  \Delta_i((k+1)T)= &\, Q\left(\frac{x_i((k+1)T)-\hat{x}_i(kT)}{\beta(kT)}\right),
  \end{aligned} \right.
\end{equation}
where $\hat{x}_i$ is the estimate of the state of agent $i$ by agent $i$ itself and its neighbors. In (\ref{eq124}), the scaling function $\beta(kT)$ is updated by (\ref{eq100}) with
\begin{equation}\label{eq125}
 \gamma=\left\{
  \begin{aligned}
  \gamma_1,  ~\textrm{if}~ k+1 \in H_v,  \\
  \gamma_2,  ~\textrm{if}~ k+1 \notin H_v,
  \end{aligned} \right.
\end{equation}
where $0<\gamma_1<1$ and $\gamma_2>1$. Accordingly, the control protocol in \cite{Feng-2020b} is given by
\begin{equation}\label{eq130}
 u_i(kT)=  K_c\sum_{j\in\mathcal{N}_i}a_{ij}\left(\hat{x}_j(kT)-\hat{x}_i(kT)\right),
 \end{equation}
where the feedback gain $K_c$ is selected such that the eigenvalues of the matrix $\textrm{diag}\{A-\lambda_2BK_c, \ldots, A-\lambda_NBK_c\}$ locate inside the unit circle.

In the sequel, several remarks are presented that provide  intuitive explanations on the properties and advantages of the approach developed in this paper.

\emph{Remark 5 (Design Complexity):} Compared with \cite{Feng-2020b}, the control design in this paper is relatively simpler.  When the system is subject to  DoS attacks, $\mathcal{N}^{\textrm{DoS}}_i(kT)=\emptyset$, and hence by (\ref{eq103}) and (\ref{eq101}), one gets (\ref{eq110}), i.e., $h_i((k+1)T) =  h_i(kT)$.  This implies that the variable $h_i(kT)$ keeps invariant under DoS attacks. Consequently, in (\ref{linear encoder}), we are able to make the state of the encoder invariant during DoS attacks, and hence the parameter of the scaling function $\gamma=1$ if the transmission fails. By contrast, in (\ref{eq124}), the control protocol is designed as (\ref{eq130}) in which the state estimation under DoS in the encoder is calculated according to the open-loop equation $\hat{x}_i((k+1)T)=A\hat{x}_i(kT)$. In this case, the consensus error driven by the control protocol (\ref{eq130}) may diverge during  DoS attacks \cite{Feng-2020b}. Therefore, in \cite{Feng-2020b}, the scaling function $\beta(kT)$ must be updated accordingly to make sure that the quantizer is unsaturated during DoS attacks. That is, zoom-out the term $\left[x_i((k+1)T)-\hat{x}_i(kT)\right]$ by introducing an extra parameter $\gamma_2$. It should be mentioned that the calculation of $\gamma_2$ is not an easy work. In fact, the value of $\gamma_2$ is decided by the system matrices $A$ and $B$, the Laplacian matrix $\mathcal{L}$, and $\Omega$ which represents the maximum number of consecutive packet losses induced by DoS attacks. Also note that the information of $\Omega$ is not required for controller design in this paper. \IEEEQED

\emph{Remark 6 (Consensus Performance):} The consensus performance can be evaluated by the convergence speed, transient performance, and final consensus value. For the convergence speed and transient performance, it is mainly decided by the scaling function $\beta(kT)$. During DoS attacks, the consensus error in \cite{Feng-2020b} may diverge while in this paper it keeps invariant (see (\ref{eq112})). In this sense, the approach developed in this paper is able to lead to faster convergence speed and better transient performance under the comparable level of DoS attacks. For the final consensus value, in \cite{Feng-2020b}, it diverges to infinity when $A$ has at least one eigenvalue outside the unit circle. On the contrary, the final consensus value in Theorem 3 is given in terms of the initial condition of the multi-agent system. This is important for some practical applications, such as distributed computation \cite{Kuhn-2010}. \IEEEQED

\emph{Remark 7 (Required Data Rate):} Compared with \cite{Feng-2020b}, the approach developed in this paper is capable of achieving lower data rate. On one hand, we encode the information of a linear combination of the agent state (i.e., $h_i$) rather than the agent state $x_i$ itself as in (\ref{eq124}). That is to say, the dimension of the encoding-decoding channel between each pair of adjacent agents in this paper is 1 while in \cite{Feng-2020b} is $r$ which is the dimension of $x_i$. On the other hand, we show that, no matter how many agents there are, our proposed protocol guarantees consensus with a fixed number of quantization level. More importantly, the lowest number of the quantization level we achieve is 3 (i.e., $K=1$, a two-bit quantizer). In \cite{Feng-2020b}, the required quantization level is proportional to the value of $\sqrt{Nr}$. This limits its capability for a large group of agents. \IEEEQED

\emph{Remark 8 (DoS Attack Tolerance):} For the zooming-in and zooming-out mechanism developed in \cite{Feng-2020b}, as the consensus error may diverge during DoS attacks, the level of tolerable DoS attacks is expected to be low. More specifically, the successful transmissions must occupy sufficient proportion of the total transmissions, to compensate for the increase of the consensus error caused by those unsuccessful transmissions. In \cite{Feng-2020b}, the tolerable DoS level is given by $\frac{1}{T_D}+\frac{T}{\tau_D}<\frac{-\ln \gamma_1}{\ln \gamma_2 - \ln \gamma_1}$. Note that the term $\frac{-\ln \gamma_1}{\ln \gamma_2 - \ln \gamma_1}$ can be much smaller than 1 in case the resulting $\gamma_2$ is large, which implies a lower tolerance to DoS. In this paper, we only require that the number of consecutive packet losses induced by DoS attacks is upper bounded, that is, $\frac{1}{T_D}+\frac{T}{\tau_D}<1$.  As discussed in \cite{Feng-2017}, the bound ``$<1$" is the best bound that one can achieve in the sense that if the bound is violated, a DoS attacker may corrupt all the transmission attempts and consensus is not possible by any controller design. From this perspective, the control scheme in this paper largely increases the tolerance to DoS attacks with respect to \cite{Feng-2020b}.
\IEEEQED

\section{Examples}\label{Sec_exp}

In this section, we present two simulation examples: a nonlinear academic one, demonstrating the full capabilities of the proposed control scheme, and a linear comparative example with the recent paper \cite{Feng-2020b}.

\subsection{Nonlinear Example}

\begin{figure}[!t]
  \centering
  \includegraphics[width=0.20\textwidth,bb=15 15 160 100, clip]{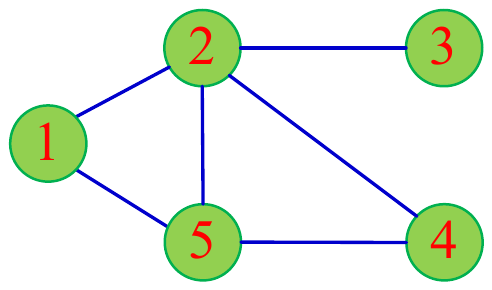}
  \caption{The digital communication network of the five agents.}\label{Network}
\end{figure}

Consider a group of five agents with the following dynamics:
\begin{equation}\label{eq18}
   \left\{
  \begin{aligned}
           \dot{x}_{i1}= & x_{i1}+x_{i2},  \\
           \dot{x}_{i2}= & x_{i3}, \\
           \dot{x}_{i3}= & -p_{i3}x_{i3}-p_{i1}x_{i2}-p_{i2}x_{i2}^3+u_i+z_i+\omega_i,\\
           \dot{z}_i =   & -(x^2_{i3}+\omega_i^2)z_i,\\  y_i= & x_{i1}, ~1\leq i\leq 5,\\
        \end{aligned} \right.
\end{equation}
where $p_{i1}$, $p_{i2}$, and $p_{i3}$ are system parameters, and $\omega_i$ is the external disturbance. System (\ref{eq18}) can be regarded as a connection of a nonlinear Duffing equation  \cite{Dong-2018} with stable zero dynamics and a first-order linear system.  Let $\rho_{i1}=x_{i1}$, $\rho_{i2}=x_{i1}+x_{i2}$, and $\rho_{i3}=x_{i1}+x_{i2}+x_{i3}$. Then system (\ref{eq18}) can be written as
\begin{equation}\label{eq26}
   \left\{
  \begin{aligned}
           \dot{\rho}_{i1}= & \rho_{i2},  \\
           \rho_{i2}= & \rho_{i3}, \\
           \dot{\rho}_{i3}= & \rho_{i3}-p_{i3}(\rho_{i3}-\rho_{i2})-p_{i1}(\rho_{i2}-\rho_{i1}) \\
           & -p_{i2}(\rho_{i2}-\rho_{i1})^3+u_i+z_i+\omega_i,\\
           \dot{z}_i =   & -\left((\rho_{i3}-\rho_{i2})^2+\omega_i^2\right)z_i, \\
          y_i=& \rho_{i1}, ~1\leq i\leq 5.\\
        \end{aligned} \right.
\end{equation}
The digital communication network among the five agents is depicted in Fig. \ref{Network}. In this simulation study, we set $p_{i1}=-1.1-0.2i$, $p_{i2}=1+0.2i$, $p_{i3}=0.4+0.1i$, and $\omega_i=\frac{1}{2i}\sin(it)$, $1\leq i\leq 5$. The initial conditions of the agents $\rho_{ij}(0)$ and $z_i(0)$, $1\leq i\leq 5$, $1\leq j\leq 3$, are randomly located in $[0, 5]$. For the proposed control scheme, let $k_1=k_2=4$, and $T=0.05$; the bounds are selected as $M_1=10$, $M_2=10$, $M_3=10$, $M_4=100$, and $C_h=45$; the ESOs are designed with $[l_{1} ~l_{2} ~l_{3} ~l_{4}]^{\rm{T}}=[4 ~6 ~4 ~1]^{\rm{T}}$, $1\leq i\leq 5$.

\begin{figure}[!t]
        \centering
        \begin{subfigure}[b]{0.24\textwidth}
            \centering
            \includegraphics[width=\textwidth,bb=20 0 320 240, clip]{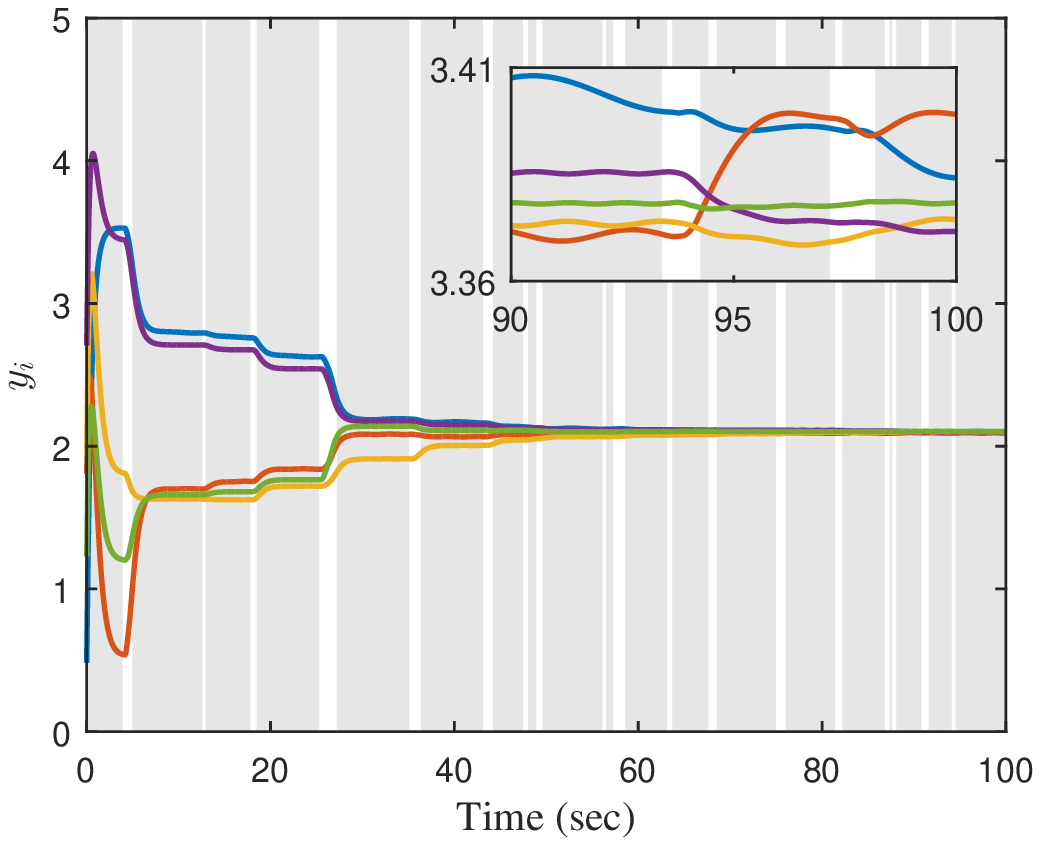}
          \caption[Fig2a]{{$\varepsilon=0.01$.}}
            \label{fig2a}
        \end{subfigure}
        \begin{subfigure}[b]{0.24\textwidth}
            \centering
            \includegraphics[width=\textwidth,bb=20 0 320 250, clip]{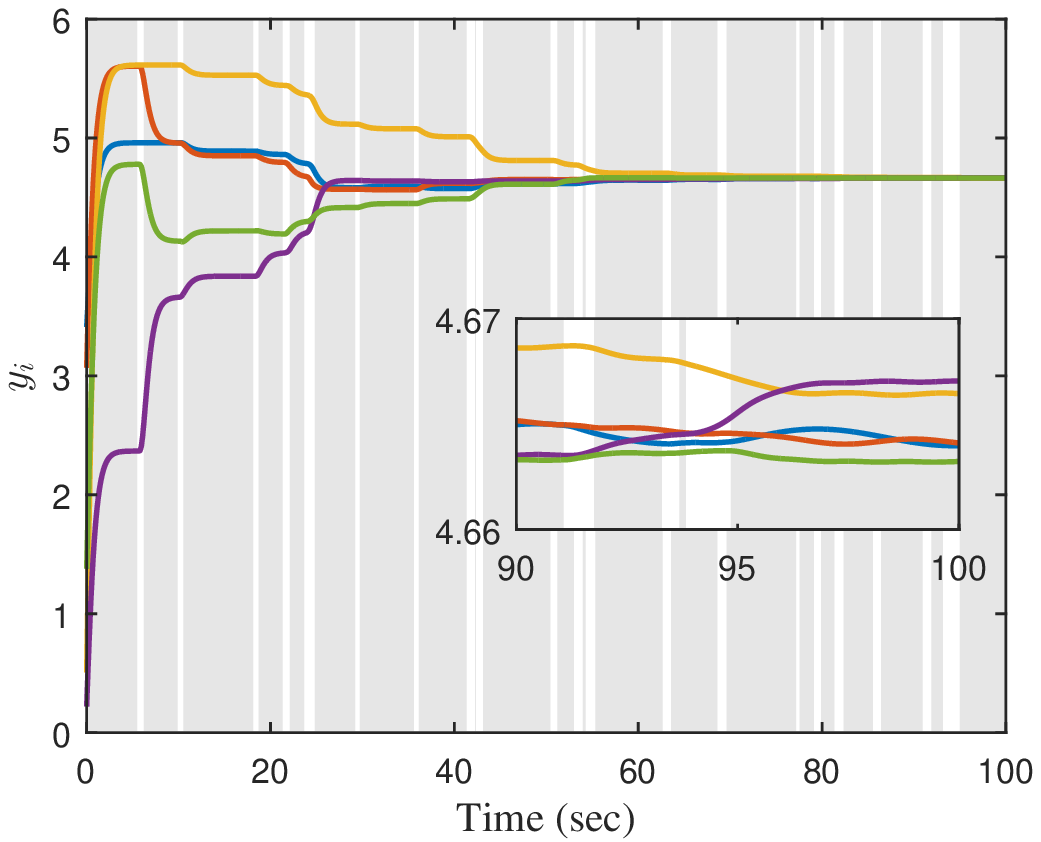}
            \caption[Fig2b]{{$\varepsilon=0.001$.}}
            \label{fig2b}
        \end{subfigure}
        \caption[Fig2]{Trajectories of $y_i$ with $K=10$.  The DoS attacks (colored in gray) are present for $\sim$90\% of total time.}
        \label{fig2}
    \end{figure}

\begin{figure}[!t]
 \centering
 \includegraphics[width=0.40\textwidth,bb=20 0 390 320, clip]{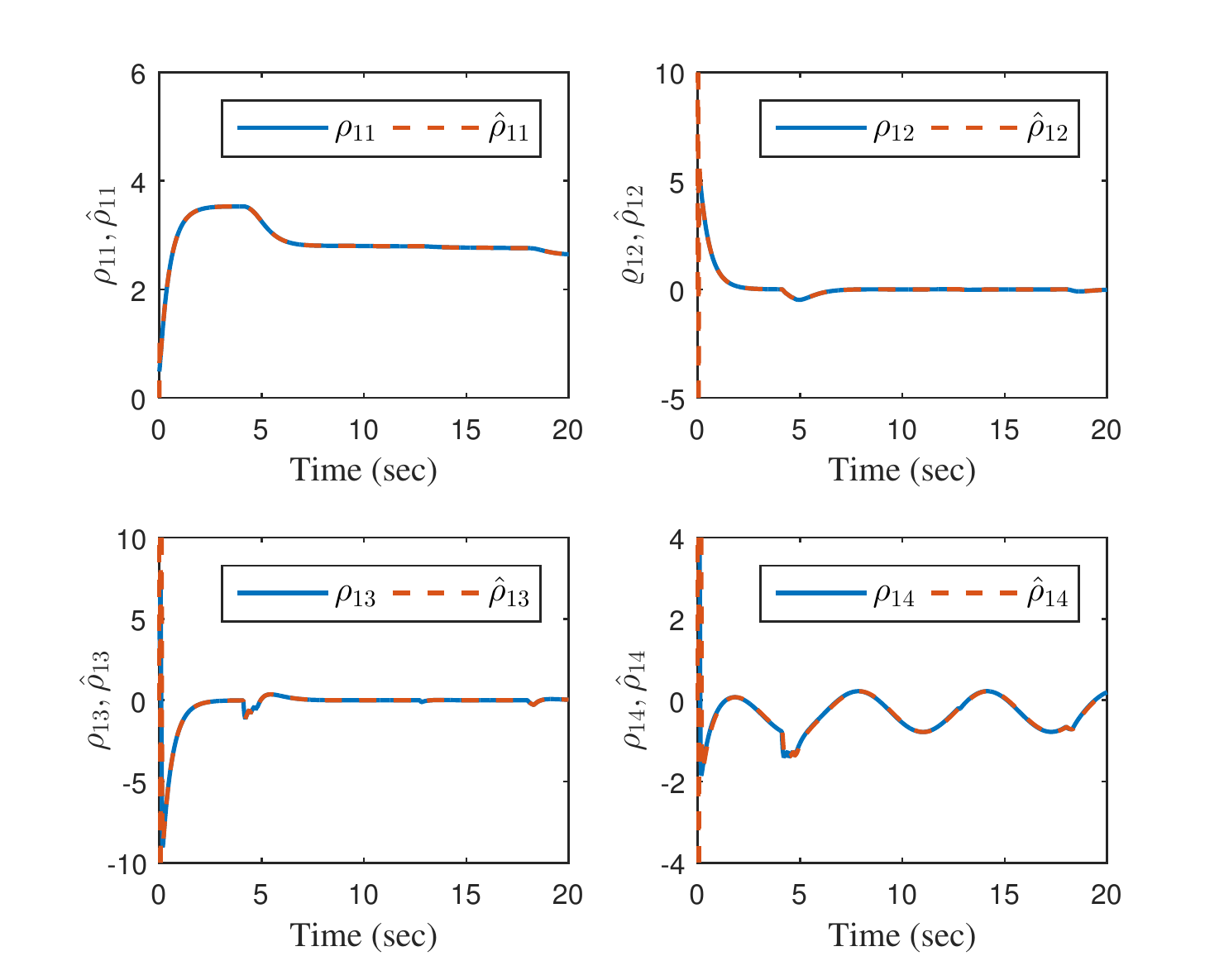}
 \caption{Performance of the ESO with $\varepsilon=0.01$.}\label{fig3}
\end{figure}

\subsubsection{Simulation with High Quantization Level} First, we simulate the high quantization level case. According to Theorem 1, let $c=1$, $\gamma_1=0.96$, $\beta_0=10$, and $K=10$. We investigate a randomly generated sustained DoS. Here for simulation purpose, the maximal duration of the DoS is limited by 9 seconds, i.e., the number of consecutive packet losses is no larger than 180, and the DoS attacks are present for $\sim$90\% of total time. Fig. \ref{fig2} shows the simulation results with $\varepsilon=0.01$ and $\varepsilon=0.001$. From this figure, one can observe that the multi-agent system with unknown nonlinear dynamics, data-rate constraints, and DoS attacks achieved practical convergence. Also note that smaller $\varepsilon$ leads to smaller residual consensus error. Fig. \ref{fig3} depicts the response of the ESO for agent 1 with $\varepsilon=0.01$. One can see that both the agent states $\rho_{1j}$, $1\leq j\leq 3$, and the extended state $\rho_{14}\triangleq \rho_{13}-p_{13}(\rho_{13}-\rho_{12})-p_{11}(\rho_{12}-\rho_{11})-p_{12}(\rho_{12}-\rho_{11})^3+z_1+\omega_1$ are well-estimated.

\subsubsection{Simulation with Low Quantization Level} 
\begin{figure}[!t]
 \centering
 \includegraphics[width=0.48\textwidth,bb=70 0 710 215, clip]{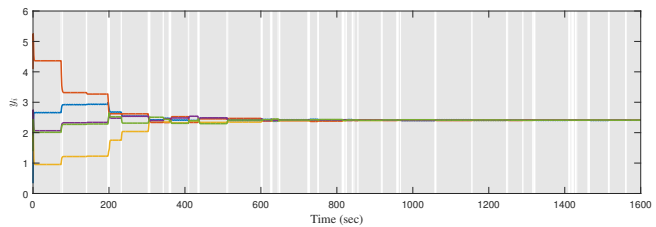}
 \caption{Trajectories of $y_i$ with $K=1$ and $\sim$95\% DoS attacks.}\label{fig4}
\end{figure}

Next, we investigate the performance of the proposed control protocol with two-bit quantizers (i.e., $K=1$). According to the conditions in Theorem 2, we select $\epsilon_0=0.9$, $K=1$, $c=0.5$, $\gamma_1=0.9975$, and $\beta_0=35$. Here we simulate a more serious DoS scenario. The duration of each DoS attack is randomly selected between 0 to 95 seconds, while the duration without DoS attack is randomly selected between 0 to 5 seconds. That is, the DoS attacks are present for $\sim$95\% of total time. Fig. \ref{fig4} shows the trajectories of $y_i$, $1\leq i\leq 5$, with $K=1$ and  $\sim$95\% DoS attacks. One can see that the agents achieved output consensus with serious DoS attacks and merely two-bit information exchange between each pair of adjacent agents at each time step. The simulation results are consistent with the theoretical results in Theorem 2.

\subsection{Comparative Example}

We consider the diving consensus problem of five autonomous underwater vehicles (AUVs). The discrete-time  model of the $i$th AUV with sampling period 1 second is as follows \cite{Rez-2021} (we have transformed the system into controllable canonical form):
\begin{align}\label{eq131}
  \left[
    \begin{array}{c}
      x_{i1}(k+1) \\
      x_{i2}(k+1) \\
      x_{i3}(k+1) \\
    \end{array}
  \right]=& \left[\begin{array}{ccc}
                     0 & 1 & 0 \\
                     0 & 0 & 1 \\
                     0.216 & -1.502 & 2.286 \\
                   \end{array}
                 \right]\left[
    \begin{array}{c}
      x_{i1}(k) \\
      x_{i2}(k) \\
      x_{i3}(k) \\
    \end{array}
  \right] \nonumber \\
 & +\left[
    \begin{array}{c}
     0 \\
     0 \\
     1 \\
    \end{array}
  \right] u_i(k), ~1\leq i\leq 5.
\end{align}
The communication graph of the AUVs is also given by Fig. \ref{Network}. The initial conditions are set as  $x_1(0)=[57.55 ~181.56~ 113.06]^{\textrm{T}}$, $x_2(0)=[96.21 ~32.08$ $180.69]^{\textrm{T}}$, $x_3(0)=[26.76 ~75.41 ~197.66]^{\textrm{T}}$, $x_4(0)=[71.21 ~118.11$ $79.59]^{\textrm{T}}$, and $x_5(0)=[108.91$ $50.44 ~19.41]^{\textrm{T}}$, which are generated by the \emph{rand} function in \emph{Matlab}  with $\|x_i(0)\|_{\infty}\leq 200$.

\subsubsection{Simulation with Low-Level DoS Attacks} 
\begin{figure}[!t]
        \centering
        \begin{subfigure}[b]{0.24\textwidth}
          \centering
          \includegraphics[width=\textwidth,bb=0 0 325 240, clip]{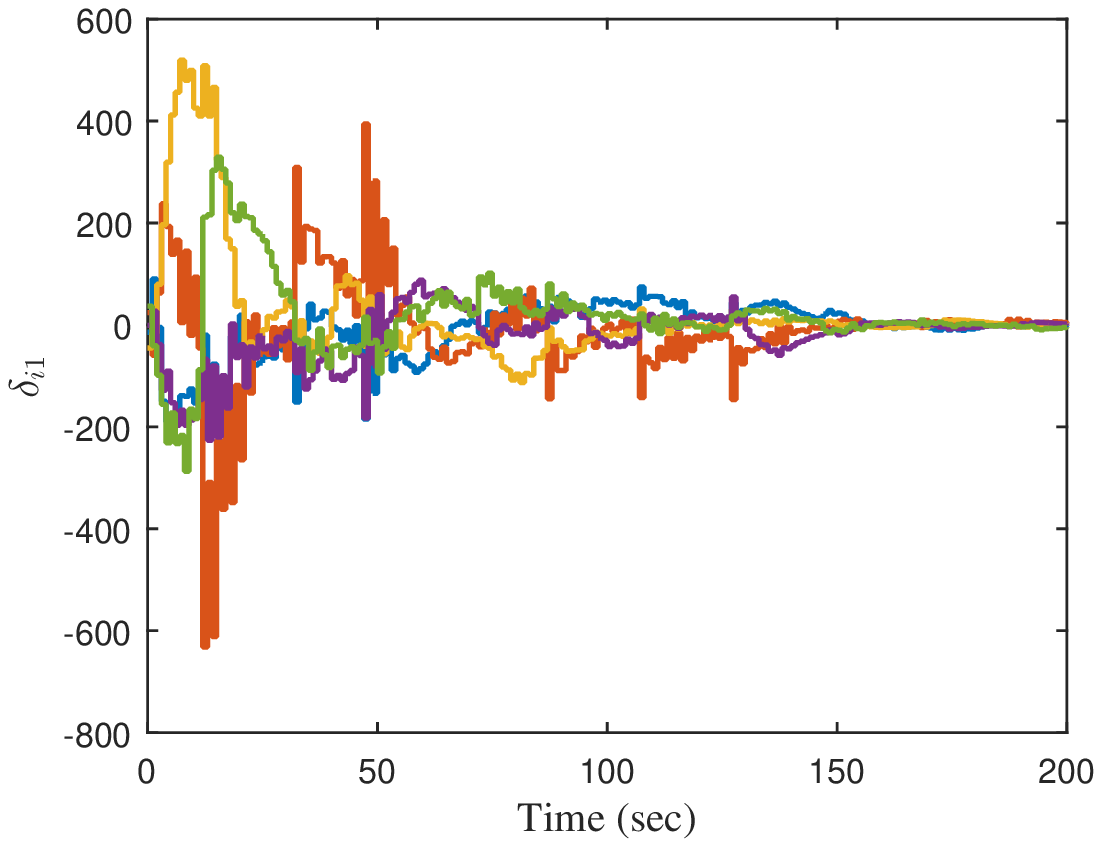}
          \caption{Control protocol in \cite{Feng-2020b}.}\label{fig5a}
        \end{subfigure}
        \begin{subfigure}[b]{0.24\textwidth}
         \centering
         \includegraphics[width=\textwidth,bb=0 0 325 245, clip]{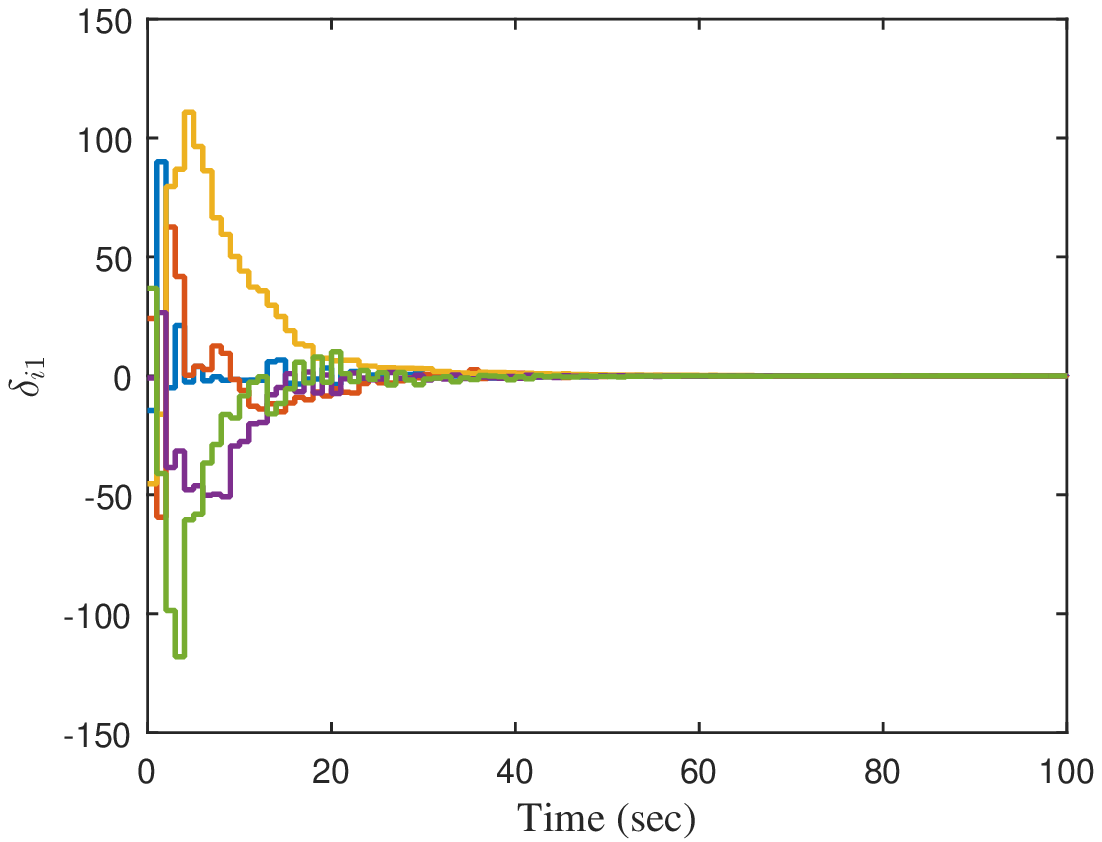}
         \caption{Control protocol in this paper.}\label{fig5b}
        \end{subfigure}
  \caption[Fig6]{Responses of $\delta_{i1}(kT)$ under the control protocols in \cite{Feng-2020b} and this paper with $K=1600$ and $\sim5\%$ DoS attacks.} \label{fig5}
    \end{figure}

First, we simulate the low-level DoS attacks case. For the control protocol in \cite{Feng-2020b}, let $K_c=[0.0720 ~-0.4426 ~0.4029]$, $\gamma_1=0.9$, $\gamma_2=4.8$, $\beta_0=100$, and $K=1600$ (see Theorem 1 in \cite{Feng-2020b} for the procedure of the calculation of these parameters). For the control protocol in this paper, according to Theorem 3, we select $k_1=-0.25$, $k_2=0$, $C_h=150$, $\epsilon_0=0.5$, $c=0.2$, $\gamma_1=0.9$, $\beta_0=100$, and $K=1600$. Note that the same quantization parameter $K$, initial value of the scaling function $\beta_0$, and parameter $\gamma_1$ are adopted to make a fair comparison. The low-level DoS attacks are present for $\sim5\%$ of total time. Fig. \ref{fig5} shows the responses of $\delta_{i1}(kT)=x_{i1}(kT)-\frac{1}{5}\sum_{j=1}^5x_{j1}(kT)$, $1\leq i\leq 5$, under the two control protocols with transmissions failed at $k=10, 30, 45, 70, 85, 105, 125, 145, 165, 185$. From this figure, one can observe that the control protocol in this paper achieves much better consensus performance. The main reason leads to this improvement is that the consensus error diverges during the DoS attacks under the control protocol in \cite{Feng-2020b}, while is held unchanged in this paper.

\subsubsection{Simulation with High-Level DoS Attacks} 
\begin{figure}[!t]
        \centering
        \begin{subfigure}[b]{0.24\textwidth}
        \centering
        \includegraphics[width=\textwidth,bb=0 0 325 240, clip]{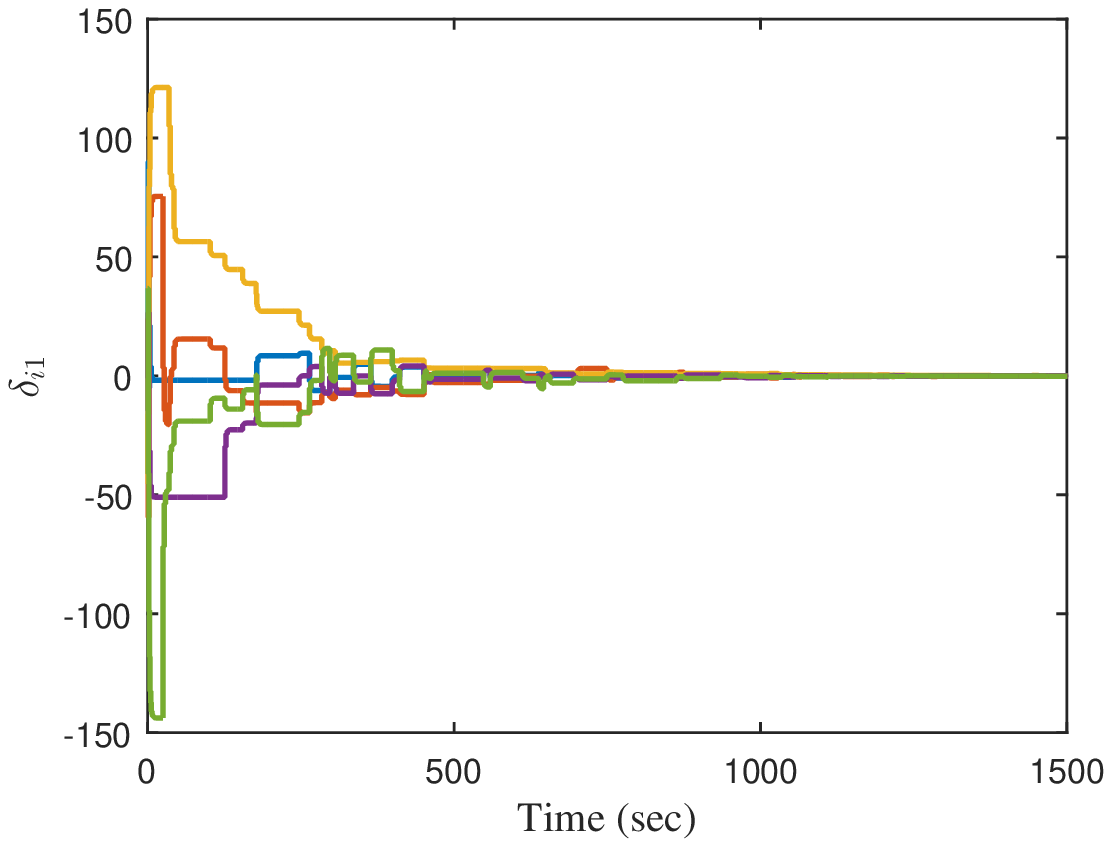}
        \caption{$K=1600$.}\label{fig6a}
        \end{subfigure}
        \begin{subfigure}[b]{0.24\textwidth}
        \centering
        \includegraphics[width=\textwidth,bb=0 0 325 245, clip]{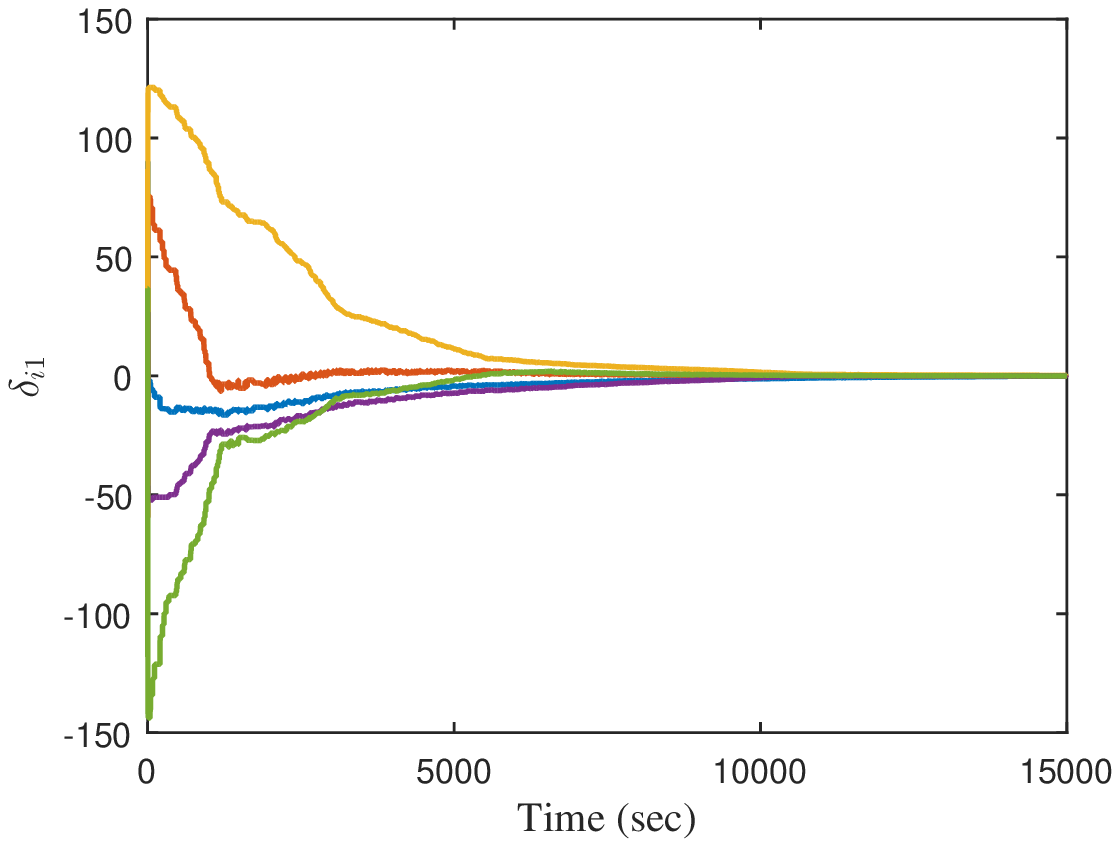}
        \caption{$K=1$.}\label{fig6b}
        \end{subfigure}
  \caption[Fig6]{Responses of $\delta_{i1}(kT)$ under the proposed control protocol with different quantization levels and $\sim95\%$ DoS attacks.} \label{fig6}
\end{figure}

We further simulate the high-level DoS attacks case, i.e., the DoS attacks are present for $\sim95\%$ of total time. For the control protocol in \cite{Feng-2020b}, the consensus cannot be achieved if the level of DoS attacks increases to $\sim 0.07$. In contrast to \cite{Feng-2020b}, the control protocol in this paper can tolerate arbitrarily serious DoS attacks as long as their induced maximal number of consecutive
packet losses is bounded (i.e., present less than $100\%$ of total time). Fig. \ref{fig6a} illustrates the responses of $\delta_{i1}(kT)$ under the proposed control protocol with same parameters as in the previous simulation.

Now we show the potentials of data rate reduction by the control scheme developed in this paper. For the multi-agent system (\ref{eq131}), the number of quantization levels in \cite{Feng-2020b} must satisfy $2K+1\geq 2711$, while in this paper $2K+1\geq 3$. According to Theorem 3, we now select $K=1$, $\gamma_1=0.9925$, and $\beta_0=60$. The responses of $\delta_{i1}(kT)$ with 3-level quantizers and $\sim95\%$ DoS attacks are depicted in Fig. \ref{fig6b}, from which one can see that the consensus has been successfully achieved.

Finally, we mention that the final consensus value of our proposed control protocol in the simulations is $x_i=[133.40 ~133.40 ~133.40]^{\rm{T}}$, which is consistent with the theoretical computation (\ref{eq127}), and independent with the quantization parameter $K$ and DoS attacks. On the contrary, the AUVs' states under the control protocol in \cite{Feng-2020b} trend  to infinity, as the state matrix in (\ref{eq131}) is unstable.

\section{Conclusions}\label{Sec_con}

This paper investigated the quantized consensus problem for uncertain nonlinear multi-agent systems under data-rate constraints and DoS attacks. An ESO-based dynamic encoding-decoding scheme with zooming-in and holding capabilities was developed, which exhibits several intriguing properties such as better consensus performance, lower data-rate requirement, and higher tolerance to DoS attacks. In particular, it was shown that the quantized consensus can be achieved under two-bit data rate and any DoS attacks inducing bounded consecutive packet losses. Two numerical examples verified the capabilities and superiorities of the developed control scheme.

In this paper, as a preliminary theoretical research, we made some necessary assumptions.  Future research efforts will be devoted towards extending the developed control scheme to more challenging scenarios, such as directed networks \cite{Lin-2013}, with network transmission delays \cite{Liu-2011},  unreliable or no acknowledgement signal \cite{Liu-2021b,Lin-2019},  and DoS attacks that  affect only a part of the communication graph.

\section*{Appendix: Proof of Theorem 1}\label{AppenA}

To facilitate the analysis, let $\alpha(kT)= \delta(kT)/\beta(kT)$ and $\zeta(kT)=\mu(kT)/\beta(kT)$. Before stating the proof of Theorem 1, we need two additional propositions. In the following, Proposition 1 is to analyze the convergence of the ESO, and to verify that the acculturated ESO estimation error will not lead to quantizer saturation if we let $\beta(kT)\rightarrow \sqrt{\varepsilon}$. Proposition 2 is to show the  boundedness of $\alpha(v_iT)$.

\emph{Proposition 1:} Consider the system (\ref{eq1}) with ESO (\ref{eq10}). Suppose Assumptions A2 and A3 are satisfied, and the agents' states $(\rho_i, z_i)$ are bounded. Let $M_{j}\geq \sup_{1\leq i\leq N}|\rho_{ij}|$, $1\leq j\leq r+1$. Then there exists $\varepsilon^*>0$ such that $\forall \varepsilon\in(0,\varepsilon^*)$:
\begin{itemize}
\item there exists $\tau_0(\varepsilon)>0$ satisfying $\lim_{\varepsilon\rightarrow 0}\tau_0(\varepsilon)=0$, such that $\forall t\in[\tau_0(\varepsilon), \infty)$,
\begin{equation}\label{eq72}
    |\rho_{ij}(t)-\hat{\rho}_{ij}(t)|=O(\varepsilon^{r+2-j}), ~1\leq j\leq r+1.
\end{equation}
\item for any time interval $[\tau_1, \tau_2]\subseteq [0,\infty)$,
\begin{equation}\label{eq73}
 \frac{\int_{\tau_1}^{\tau_2}\left(\rho_{ij}(\tau)-\overline{\rho}_{ij}(\tau)\right)\textrm{d}\tau}{\sqrt{\varepsilon}}\rightarrow 0 ~\textrm{as}~ \varepsilon\rightarrow 0, ~1\leq j\leq r+1.
\end{equation}
\end{itemize}

\emph{Proof of Proposition 1:} Consider the scaled ESO estimation error $\eta_i=[\eta_{i1}, \ldots, \eta_{i,r+1}]^{\textrm{T}}\in\mathbb{R}^{r+1}$ with $\eta_{ij}=\frac{\rho_{ij}-\hat{\rho}_{ij}}{\varepsilon^{r+1-j}}$, $1\leq j\leq r+1$. By (\ref{eq32}) and (\ref{eq10}), one has
\begin{equation}\label{eq74}
\dot{\eta}_i=\frac{1}{\varepsilon}E\eta_i+\bar{B}\dot{F}_i(\rho_i,z_i,\omega_i),
\end{equation}
where $E$ is given by (\ref{eq34}), $\bar{B}=[0, B_{\rho}^{\textrm{T}}]^{\textrm{T}}$, and
\begin{align*}
\dot{F}_{i}(\rho_i,z_i, \omega_i)=& \sum_{j=1}^{r-1}\rho_{i,j+1}\frac{\partial F_i}{\partial \rho_{ij}}+\left[F_{i}(\rho_i,z_i, \omega_i)+u_i\right]\frac{\partial F_i}{\partial \rho_{ir}} \nonumber \\
 & +F_{i0}(\rho_i,z_i, \omega_i)\frac{\partial F_i}{\partial z_{i}}+\dot{\omega}_i\frac{\partial F_i}{\partial \omega_{i}}.
\end{align*}
By the boundedness of $\rho_i$, and Assumptions A2 and A3, there exists $\varepsilon$-independent positive constant $\iota_{1i}$ such that $|\dot{F}_i(\rho_i,z_i,\omega_i)|\leq \iota_{1i}$. Recall that the matrix $E$ is Hurwitz by design. One can
select a Lyapunov function candidate $V_i(\eta_i)=\eta_i^{\rm{T}}P\eta_i$, where $P\in\mathbb{R}^{(r+1)\times (r+1)}$ is the unique positive definite matrix solution to the equation $PE+E^{\rm{T}}P=-I_{r+1}$.  Let $\sigma_{1}$ and $\sigma_{2}$ be the minimal and maximal eigenvalues of the matrix $P$, respectively. It follows that $\sigma_{1}\|\eta_i\|^2 \leq  V_{i}(\eta_i)\leq \sigma_{2}\|\eta_i\|^2$. Then, by (\ref{eq74}), one has
\begin{align}\label{eq75}
\frac{\textrm{d}V_i(\eta_i)}{\textrm{d} t}= & \frac{1}{\varepsilon}\left(E^{\textrm{T}}P+PE\right)\|\eta_i\|^2  \nonumber\\
& +\left(\dot{F}^{\textrm{T}}_i(\rho_i,z_i,\omega_i)\bar{B}^{\textrm{T}}P\eta_i+\eta_i^{\rm{T}}P\bar{B}\dot{F}_i(\rho_i,z_i,\omega_i)\right) \nonumber \\
\leq & -\frac{1}{\varepsilon}\|\eta_i\|^2+2\sigma_{2}\iota_{1i} \|\eta_i\|  \nonumber \\
\leq & -\frac{1}{\sigma_{2}\varepsilon}V_i(\eta_i)+\frac{2\sigma_{2}\iota_{1i}}{\sqrt{\sigma_{1}}}\sqrt{V_i(\eta_i)}.
\end{align}
By $\frac{\textrm{d}V_i(\eta_i)}{\textrm{d}t}=2\sqrt{V_i(\eta_i)}\frac{\textrm{d}\sqrt{V_i(\eta_i)}}{\textrm{d}t}$, the theory of ordinary differential inequality, and some straightforward manipulations, one can obtain
\begin{align}\label{eq76}
 \|\eta_i\| \leq &  \frac{\sqrt{V_i(\eta_i)}}{\sqrt{\sigma_{1}}} \nonumber \\
 \leq &  \left(\frac{\sqrt{V_i(\eta_i(0))}}{\sqrt{\sigma_{1}}}-\frac{2\sigma_{2}^2\iota_{1i}\varepsilon}{\sigma_{1}}\right)e^{-\frac{1}{2\sigma_{2}\varepsilon}t} +\frac{2\sigma_{2}^2\iota_{1i}\varepsilon}{\sigma_{1}} \nonumber \\
 \leq &\sqrt{\frac{\sigma_{2}}{\sigma_{1}}}\|\eta_i(0)\|e^{-\frac{1}{2\sigma_2\varepsilon}t}+\frac{2\sigma_{2}^2\iota_{1i}\varepsilon}{\sigma_{1}}.
\end{align}
Note that the right hand side of (\ref{eq76}) is of the order of $O(\varepsilon)$ for all $t\geq \tau_0(\varepsilon)\triangleq -2(r+1)\sigma_2\varepsilon \ln \varepsilon$. Since $\tau_0(\varepsilon)\rightarrow 0$ uniformly as $\varepsilon\rightarrow 0$, and  $|\rho_{ij}-\hat{\rho}_{ij}|=\varepsilon^{r+1-j}|\eta_{ij}|$, $1\leq j\leq r+1$, the conclusion in the first bullet holds. What is more, the saturations will not be invoked after the transient period of the observer, i.e., $\overline{\rho}_{ij}=\hat{\rho}_{ij}$, $t\geq \tau_0(\varepsilon)$.

For the second bullet, consider three cases.

\emph{Case 1):} $\tau_2\leq \tau_0(\varepsilon)$. In this case, one has
\begin{align}\label{eq150}
& \frac{\int_{\tau_1}^{\tau_2}\left(\rho_{ij}(\tau)-\overline{\rho}_{ij}(\tau)\right)\textrm{d}\tau}{\sqrt{\varepsilon}} \nonumber \\
 \leq & 2M_{j}\frac{\tau_0(\varepsilon)}{\sqrt{\varepsilon}} \rightarrow 0 ~\textrm{as}~ \varepsilon \rightarrow 0, ~1\leq j\leq r+1.
\end{align}

\emph{Case 2):} $\tau_1\geq \tau_0(\varepsilon)$. In this case, we start with $j=1$. By (\ref{eq74}), one has $\rho_{i1}-\hat{\rho}_{i1}=\frac{\varepsilon^{r+1}}{l_{r+1}}\left(\dot{F}_i(x_i,z_i,\omega_i)-\dot{\eta}_{i,r+1}\right)$. It follows that
\begin{align}\label{eq77}
& \frac{\int_{\tau_1}^{\tau_2}\left(\rho_{i1}(\tau)-\overline{\rho}_{i1}(\tau)\right)\textrm{d}\tau}{\sqrt{\varepsilon}}  \nonumber \\
= & \frac{\varepsilon^{r+\frac{1}{2}}}{l_{r+1}} \left[\left(F_i(x_i,z_i,\omega_i)-\eta_{i,r+1}\right)|_{t=\tau_2}\right. \nonumber \\
&  \left. -\left(F_i(x_i,z_i,\omega_i)-\eta_{i,r+1}\right)|_{t=\tau_1}\right].
\end{align}
Since $F_i(x_i,z_i,\omega_i)$ is bounded, and $\eta_{i,r+1}=O(\varepsilon)$, $\forall t\geq \tau_0(\varepsilon)$, the right hand side of (\ref{eq77}) is of the order of $O(\varepsilon^{r+\frac{1}{2}})$
%
%
For $j=2$, by (\ref{eq74}), one has $\rho_{i2}-\hat{\rho}_{i2}=\varepsilon^{r}\dot{\eta}_{i1}+\frac{l_{i1}}{\varepsilon}(\rho_{i1}-\hat{\rho}_{i1})$. It follows that
\begin{align}\label{eq78}
& \frac{\int_{\tau_1}^{\tau_2}\left(\rho_{i2}(\tau)-\overline{\rho}_{i2}(\tau)\right)\textrm{d}\tau}{\sqrt{\varepsilon}} \nonumber \\
= & \varepsilon^{r-\frac{1}{2}}(\eta_{i1}(\tau_2)-\eta_{i1}(\tau_1))+\frac{l_{i1}}{\varepsilon}\frac{\int_{\tau_1}^{\tau_2}(\rho_{i1}(\tau)-\hat{\rho}_{i1}(\tau))\textrm{d}\tau}{\sqrt{\varepsilon}}.
\end{align}
By (\ref{eq77}), the right hand side of (\ref{eq78}) is of the order of $O(\varepsilon^{r-\frac{1}{2}})$.  Then by a simple iterative procedure, one can prove that
\begin{equation}\label{eq151}
\frac{\int_{\tau_1}^{\tau_2}\left(\rho_{ij}(\tau)-\overline{\rho}_{ij}(\tau)\right)\textrm{d}\tau}{\sqrt{\varepsilon}}=O(\varepsilon^{r+\frac{3}{2}-j}),~ 1\leq j\leq r+1.
\end{equation}

\emph{Case 3):} $\tau_1< \tau_0(\varepsilon) <\tau_2$. Note that $\int_{\tau_1}^{\tau_2}\left(\cdot\right)\textrm{d}\tau= \int_{\tau_1}^{\tau_0(\varepsilon)}\left(\cdot\right)\textrm{d}\tau+\int_{\tau_0(\varepsilon)}^{\tau_2}\left(\cdot\right)\textrm{d}\tau$. Then by Cases 1) and 2), one can conclude that $\frac{\int_{\tau_1}^{\tau_2}\left(\rho_{ij}(\tau)-\overline{\rho}_{ij}(\tau)\right)\textrm{d}\tau}{\sqrt{\varepsilon}}\rightarrow 0$ as $\varepsilon\rightarrow 0$ in this case.

Finally, combining Cases 1) to 3) leads to the conclusion in (\ref{eq73}). This completes the proof of Proposition 1. \IEEEQED

\emph{Proposition 2:}
There exists $\varepsilon^{\ddag}>0$ such that $\forall \varepsilon\in(0,\varepsilon^{\ddag})$, \begin{equation}\label{eq81}
    \|\alpha(v_iT)\|<\overline{\alpha}, ~i=0,1, \ldots,
\end{equation}
where $\overline{\alpha}=\max\left\{\frac{2\sqrt{N}C_h}{\beta_0}, \frac{cT\sqrt{N}\lambda_N}{2\gamma_1(\gamma_1-\rho_h)}\right\}$. Furthermore, the quantizer will never be saturated.

\emph{Proof of Proposition 2:}  Proposition 2 will be proved recursively. For $t=v_0T=0$, by (\ref{eq28}) and the fact that $\|\nu\|_{\infty}\leq \|\nu\|\leq \sqrt{N}\|\nu\|_{\infty}$, $\forall \nu\in\mathbb{R}^N$, one has
\begin{equation}\label{eq85}
    \|\alpha(0)\|=\frac{\|\delta(0)\|}{\beta_0}\leq \frac{\sqrt{N}\|\delta(0)\|_{\infty}}{\beta_0}\leq \frac{2\sqrt{N}C_h}{\beta_0}<\overline{\alpha}.
\end{equation}
What is more, by $\overline{H}(0)=\widehat{\overline{H}}(0)=\textbf{0}_N$, one has
\begin{equation}\label{eq86}
 \|\zeta(0)\|_{\infty}=\frac{\|\overline{H}(0)-\widehat{\overline{H}}(0)\|_{\infty}}{\beta_0} =0.
\end{equation}

According to the update mechanism of the scaling function, $\beta(kT)$ will converge to $\sqrt{\varepsilon}$ at some successful transmission time instant and then keep this value afterwards. Let $v_{m^*}T$ denote the first time instant when $\beta(kT)=\sqrt{\varepsilon}$. We consider three cases, that is $v_m<v_{m^*}$, $v_m=v_{m^*}$, and $v_m>v_{m^*}$.

 \begin{figure}
   \centering
   \includegraphics[width=0.48\textwidth,bb=10 15 540 60, clip]{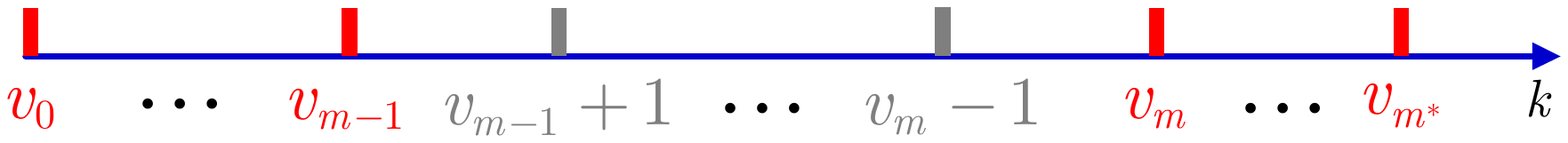}
   \caption{Sequences of the successful (colored in red) and unsuccessful (colored in gray) transmissions.}\label{fig_timeline}
 \end{figure}

\emph{Case 1):} $v_m<v_{m^*}$. Recall that $v_0T, \ldots, v_mT, \ldots$ denote the successful transmission time instants (see Fig. \ref{fig_timeline}). Let $(v_{m-1}+1)T, \ldots,$ $(v_{m}-1)T$ be the unsuccessful transmission time instants between two consecutive successful transmission time instants $v_{m-1}T$ and $v_mT$ \footnote{If $v_{m-1}+1>v_m-1$, this indicates that $v_m=v_{m-1}+1$, and there is no attack for the consecutive sampling time instants $v_{m-1}T$ and $v_mT$.}. By (\ref{eq17}), (\ref{eq88}), and $\beta((v_{m-1}+1)T)=\beta(v_{m-1}T)$, the evolution equations of $\alpha(kT)$ and $\zeta(kT)$ from $k=v_{m-1}$ to $k=v_{m-1}+1$ is given by
\begin{align}
\alpha((v_{m-1}+1)T)
=& (I-cT\mathcal{L})\alpha(v_{m-1}T)+cT\mathcal{L}\zeta(v_{m-1}T) \nonumber \\
\label{eq6}   & +\frac{\varsigma_2(v_{m-1}T)}{\beta(v_{m-1}T)},  \\
\zeta((v_{m-1}+1)T)=&(I+cT\mathcal{L})\zeta(v_{m-1}T)-cT\mathcal{L}\alpha(v_{m-1}T)\nonumber \\
\label{eq90}    & +\frac{\varsigma_3(v_{m-1}T)}{\beta(v_{m-1}T)},
\end{align}
where
\begin{align*}
 \varsigma_2(v_{m-1}T)=& cT\mathcal{L}\left[H(v_{m-1}T)-\overline{H}(v_{m-1}T)\right]  \\
 & +(I-J_N)\varsigma_1(v_{m-1}T), \\
 \varsigma_3(v_{m-1}T)=& (I+cT\mathcal{L})\left[H(v_{m-1}T)-\overline{H}(v_{m-1}T)\right] \\
 & +\varsigma_1(v_{m-1}T)  \\
 & +\overline{H}((v_{m-1}+1)T)-H((v_{m-1}+1)T).
\end{align*}

For the transmissions at time instants $(v_{m-1}+1)T, \ldots,$ $(v_{m}-2)T$, by (\ref{eq79}) and (\ref{eq88}), one has
\begin{align}
\label{eq83}    \alpha((k+1)T)=&\alpha(kT)+\frac{(I-J_N)\varsigma_{1}(kT)}{\beta(v_{m-1}T)}, \\
\label{eq94}   \zeta((k+1)T)=& \zeta(kT)+\frac{\varsigma_4(kT)}{\beta(v_{m-1}T)},
\end{align}
where $v_{m-1}+1\leq k\leq v_m-2$, and
\begin{align*}
    \varsigma_4(kT)=&\varsigma_1(kT)+H(kT)-\overline{H}(kT) \nonumber \\
    & +\overline{H}((k+1)T)-H((k+1)T).
\end{align*}
Applying a simple iteration to (\ref{eq83})-(\ref{eq94}) yields
\begin{align}\label{eq84}
\alpha((v_m-1)T) =& \alpha((v_{m-1}+1)T) \nonumber \\
& +(I-J_N)\sum_{j=v_{m-1}+1}^{v_{m}-2}\frac{\varsigma_{1}(jT)}{\beta(v_{m-1}T)},\\
\label{eq91}
\zeta((v_m-1)T)=& \zeta((v_{m-1}+1)T)+\sum_{j=v_{m-1}+1}^{v_m-2}\frac{\varsigma_4(jT)}{\beta(v_{m-1}T)}.
\end{align}

For the transmission at time instant $(v_m-1)T$, by (\ref{eq79}), (\ref{eq54}), and $\beta(v_mT)=\gamma_1\beta((v_{m}-1)T)$, one has
\begin{align}
\label{eq89}  \alpha(v_mT)=&\frac{1}{\gamma_1}\alpha((v_m-1)T)+\frac{(I-J_N)\varsigma_{1}((v_m-1)T)}{\beta(v_{m}T)}, \\
\label{eq92} \zeta(v_mT)=&\frac{1}{\gamma_1}\zeta((v_m-1)T)+\frac{\varsigma_4((v_m-1)T)}{\beta(v_mT)} \nonumber \\
    & -\frac{1}{\gamma_1}Q\left(\zeta((v_m-1)T)+\frac{\varsigma_4((v_m-1)T)}{\beta(v_{m-1}T)}\right).
\end{align}

Therefore, by (\ref{eq6})-(\ref{eq90}), (\ref{eq84})-(\ref{eq91}), and (\ref{eq89})-(\ref{eq92}), the equations that describe the evolution of $\alpha(kT)$ and $\zeta(kT)$ from $k=v_{m-1}$ to $k=v_m$ are expressed as
\begin{align}
\label{eq58} \alpha(v_mT)=& \frac{I-cT\mathcal{L}}{\gamma_1}\alpha(v_{m-1}T)+\frac{cT\mathcal{L}}{\gamma_1}\zeta(v_{m-1}T) \nonumber \\
& +\frac{\varsigma_2(v_{m-1}T)}{\beta(v_mT)}+\frac{(I-J_N)\sum_{j=v_{m-1}+1}^{v_{m}-1}\varsigma_{1}(jT)}{\beta(v_{m}T)}, \\
\label{eq93}  \zeta(v_mT) = & \frac{1}{\gamma_1}\left[\Psi(v_{m-1}T)-Q(\Psi(v_{m-1}T))\right],
\end{align}
where
\begin{align}
    \Psi(v_{m-1}T)=&(I+cT\mathcal{L})\zeta(v_{m-1}T)-cT\mathcal{L}\alpha(v_{m-1}T) \nonumber \\
\label{eq133}    &     +\frac{\varsigma_3(v_{m-1}T)}{\beta(v_{m-1}T)}+\sum_{j=v_{m-1}+1}^{v_m-1}\frac{\varsigma_4(jT)}{\beta(v_{m-1}T)}.
\end{align}

For $m=1$, according to Proposition 1, (\ref{eq28}), (\ref{eq85}), and (\ref{eq86}), one can readily verify that $\|\alpha(v_1T)\|< \overline{\alpha}$ and $\|\Psi(v_0T)\|_{\infty}< K+\frac{1}{2}$ for sufficiently small $\varepsilon$. In the following, we investigate the case that $t=v_mT$, $m\geq 2$, with the assumption that $\|\alpha(v_iT)\|<\overline{\alpha}$ and the quantizer is unsaturated for all $t=v_iT$, $1\leq i\leq m-1$.

To facilitate the analysis, let $U$ be a unitary matrix defined by $U=[\textbf{1}_N/\sqrt{N},\psi_2, \ldots, \psi_N]$, where $\psi^{\textrm{T}}_i\mathcal{L}=\lambda_i\psi_i^{\textrm{T}}$, $2\leq i\leq N$. Denote $\widetilde{\alpha}(kT)=U^{-1}\alpha(kT)=U^{\textrm{T}}\alpha(kT)$, and decompose $\widetilde{\alpha}(kT)=[\widetilde{\alpha}_1(kT), \widetilde{\alpha}_2(kT)]^{\textrm{T}}$ with a scalar $\widetilde{\alpha}_1(kT)$. One can verify that $\widetilde{\alpha}_1(kT)=0$. Let $\mathcal{M}_{\gamma_1,c}=\gamma_1^{-1}\textrm{diag}\left\{1-cT\lambda_2, \ldots, 1-cT\lambda_N\right\}$ and $\psi=[\psi_2, \ldots, \psi_N]$. It follows from (\ref{eq58}) that
\begin{align} \label{eq135}
\widetilde{\alpha}_2(v_mT)
=& \mathcal{M}_{\gamma_1,c}\widetilde{\alpha}_2(v_{m-1}T)+\frac{cT\psi^{\textrm{T}}\mathcal{L}}{\gamma_1}\zeta(v_{m-1}T) \nonumber \\
& +\frac{\psi^{\textrm{T}}\varsigma_2(v_{m-1}T)}{\beta(v_mT)}\nonumber \\
& +\frac{\psi^{\textrm{T}}(I-J_N)\sum_{j=v_{m-1}+1}^{v_{m}-1}\varsigma_{1}(jT)}{\beta(v_{m}T)}.
\end{align}
By (\ref{eq135}), one can establish the evolution equation of $\widetilde{\alpha}_2(kT)$ from $k=v_0$ to $k=v_m$. This together with $\alpha(kT)=\psi \widetilde{\alpha}_2(kT)$ and  $\widetilde{\alpha}_2(kT)=\psi^{\textrm{T}}\alpha(kT)$ yields
\begin{align}\label{eq136}
& \alpha(v_mT)=  \psi\left[\mathcal{M}_{\gamma_1,c}\right]^m\psi^{\textrm{T}}\alpha(v_0T)\nonumber \\
& \qquad +\psi\left[\mathcal{M}_{\gamma_1,c}\right]^{m-1}\frac{cT\psi^{\textrm{T}}\mathcal{L}}{\gamma_1}\zeta(v_0T)\nonumber \\
& \qquad +\psi\sum_{i=0}^{m-2}\left[\mathcal{M}_{\gamma_1,c}\right]^{i}\frac{cT\psi^{\textrm{T}}\mathcal{L}}{\gamma_1}\zeta(v_{m-1-i}T) \nonumber \\
& \qquad +\psi\sum_{i=0}^{m-1}\left[\mathcal{M}_{\gamma_1,c}\right]^{i}\frac{\psi^{\textrm{T}}\varsigma_2(v_{m-1-i}T)}{\beta(v_{m-i}T)} \nonumber \\
& \qquad +\psi\sum_{i=0}^{m-1}\left[\mathcal{M}_{\gamma_1,c}\right]^i\frac{\psi^{\textrm{T}}(I-J_N)\sum_{j=v_{m-1-i}+1}^{v_{m-i}-1}\varsigma_{1}(jT)}{\beta(v_{m-i}T)}.
\end{align}

Next we investigate the five terms on the right hand side of (\ref{eq136}) to seek for the upper-bound of $\|\alpha(v_mT)\|$. Note that $\|\mathcal{M}_{\gamma_1,c}\|\leq \frac{\rho_h}{\gamma_1}$, $\|\psi\|=1$, and $\|\mathcal{L}\|=\lambda_N$. By (\ref{eq85}), the first term satisfies
\begin{align}\label{eq65}
    \left\|\psi\left[\mathcal{M}_{\gamma_1,c}\right]^m\psi^{\textrm{T}}\alpha(v_0T)\right\|\leq &  \left(\frac{\rho_h}{\gamma_1}\right)^m\frac{2\sqrt{N}C_h}{\beta_0} \nonumber \\
    < & \left(\frac{\rho_h}{\gamma_1}\right)^{m-1}\frac{2\sqrt{N}C_h}{\beta_0}.
\end{align}
The second term in (\ref{eq136}) vanishes due to (\ref{eq86}).
For the third term, recall that we have assumed that the quantizer is unsaturated for all $t=v_iT$, $1\leq i\leq m-1$. According to (\ref{eq93}), one has $\max_{1\leq i \leq m-1}\|\zeta(v_iT)\|_{\infty}\leq \frac{1}{2\gamma_1}$, and hence
\begin{align}\label{eq66}
&  \left\|\psi\sum_{i=0}^{m-2}\left[\mathcal{M}_{\gamma_1,c}\right]^{i}\frac{cT\psi^{\textrm{T}}\mathcal{L}}{\gamma_1}\zeta(v_{m-1-i}T)  \right\| \nonumber \\
\leq & \sum_{i=0}^{m-2}\left(\frac{\rho_h}{\gamma_1}\right)^i\frac{cT\sqrt{N}\lambda_N}{2\gamma^2_1}  = \frac{cT\sqrt{N}\lambda_N}{2\gamma_1(\gamma_1-\rho_h)}\left[1-\left(\frac{\rho_h}{\gamma_1}\right)^{m-1}\right].
\end{align}
For the fourth and fifth terms, by Proposition 1 and the fact that $\frac{\rho_h}{\gamma_1}<1$, they converge to 0 as $\varepsilon\rightarrow 0$. This together with (\ref{eq28}), (\ref{eq65}), and (\ref{eq66}) yields $\|\alpha(v_mT)\|< \overline{\alpha}$. What is more, by $\|\alpha(v_{m-1}T)\|< \overline{\alpha}$, $\|\zeta(v_{m-1}T)\|_{\infty}\leq \frac{1}{2\gamma_1}$, (\ref{eq23}), and (\ref{eq28}), one has
\begin{align}\label{eq95}
& \|(I+cT\mathcal{L})\zeta(v_{m-1}T)-cT\mathcal{L}\alpha(v_{m-1}T)\|_{\infty} \nonumber \\
\leq & \|(I+cT\mathcal{L})\|_{\infty}\|\zeta(v_{m-1}T)\|_{\infty}+cT\|\mathcal{L}\|\|\alpha(v_{m-1}T)\|  \nonumber \\
< & \frac{1+2cTd^*}{2\gamma_1}+\frac{\sqrt{N}c^2T^2\lambda^2_N}{2\gamma_1(\gamma_1-\rho_h)}
=  K_1(c,\gamma_1).
\end{align}
Hence, in (\ref{eq133}), by Proposition 1 and (\ref{eq95}), one has  $\|\Psi(v_{m-1}T)\|\leq K+\frac{1}{2}$ for sufficiently small $\varepsilon$. That is, the quantizer is unsaturated when $t=v_mT$. Therefore, by induction, we conclude that $\|\alpha(v_mT)\|< \overline{\alpha}$ and the quantizer is unsaturated when $t=v_mT$, $1\leq m < m^*$.

\emph{Case 2):} $v_m=v_{m^*}$. In this case, $\beta(v_{m^*-1}T)\leq \frac{\sqrt{\varepsilon}}{\gamma_1}$ and $\beta(v_{m^*}T)=\sqrt{\varepsilon}$. By following the similar line as in (\ref{eq6})-(\ref{eq93}), one can obtain
\begin{align}
\label{eq60} \alpha(v_{m^*}T)=& \frac{\beta(v_{{m^*}-1}T)}{\sqrt{\varepsilon}} (I-cT\mathcal{L})\alpha(v_{m^*-1}T) \nonumber \\
& +\frac{\beta(v_{m^*-1}T)}{\sqrt{\varepsilon}}cT\mathcal{L}\zeta(v_{m^*-1}T) \nonumber \\
& +\frac{\varsigma_2(v_{m^*-1}T)}{\sqrt{\varepsilon}}+\frac{(I-J_N)\sum_{j=v_{m^*-1}+1}^{v_{m^*}-1}\varsigma_{1}(jT)}{\sqrt{\varepsilon}}, \\
\label{eq59}  \zeta(v_{m^*}T) = & \frac{\beta(v_{m^*-1}T)}{\sqrt{\varepsilon}}\left[\Psi(v_{m^*-1}T)-Q(\Psi(v_{m^*-1}T))\right].
\end{align}

\emph{Case 3):} $v_m>v_{m^*}$. In this case, $\beta(v_mT)=\beta(v_{m-1}T)=\sqrt{\varepsilon}$ , and hence
\begin{align}
\label{eq46}  \alpha(v_{m}T)=& (I-cT\mathcal{L})\alpha(v_{m-1}T)+cT\mathcal{L}\zeta(v_{m-1}T) \nonumber \\
    & +\frac{\varsigma_2(v_{m-1}T)}{\sqrt{\varepsilon}}+\frac{(I-J_N)\sum_{j=v_{m-1}+1}^{v_{m}-1}\varsigma_{1}(jT)}{\sqrt{\varepsilon}}, \\
\label{eq47} \zeta(v_{m}T)=& \Psi(v_{m-1}T)-Q(\Psi(v_{m-1}T)).
\end{align}
Note that in Case 2) $\frac{\beta(v_{{m^*}-1}T)}{\sqrt{\varepsilon}}\leq \frac{1}{\gamma_1}$, and in Case 3) $1<\frac{1}{\gamma_1}$. By Proposition 1 and conducting a similar analysis as in (\ref{eq135})-(\ref{eq95}), one can verify that $\|\alpha(v_iT)\|<\overline{\alpha}$ and the quantizer is unsaturated when $t=v_mT$, $m\geq m^*$.

Finally, combining Cases 1) to 3) leads to (\ref{eq81}). What is more, by Cases 1) to 3), (\ref{eq13}) and (\ref{eq5}), and the update mechanism of the encoder, one can conclude that the quantizer will never be saturated. This completes the proof of Proposition 2. \IEEEQED

The proof of Proposition 2 is inspired by \cite{Li-2011}. However, due to the uncertain nonlinear dynamics and the introduction of the DoS attacks, several new and important technique issues arise. On one hand, since the DoS destroys the information availability, the behavior of the system in the presence and absence of DoS attacks needs to be carefully analyzed. In fact, the evolution equations  (\ref{eq84}) and (\ref{eq89}) clearly reveal the properties of the proposed zooming-in and holding approach. On the other hand, in this work, the agent uncertain nonlinear  dynamics, the DoS affected quantization process, and the consensus process are highly coupled.   In (\ref{eq136}), we show that the evolution of the scaled consensus error is governed  by the accumulated ESO estimation error, the accumulated error induced by quantization, and the behavior of the DoS attacker.

Finally, we are ready to state the proof of Theorem 1.

\emph{Proof of Theorem 1:} By Propositions 1 and 2, one can readily conclude the statement in the first bullet of Theorem 1. For the second bullet, in Proposition 2, we have proved that the quantizer will never be saturated for all transmissions. According to the definition of $\alpha(kT)$ and the update mechanism of the scaling function $\beta(kT)$, for $m\geq m^*$, one has $\|\delta(v_mT)\|=\|(I-J_N)H(v_mT)\|=\beta(v_mT)\|\alpha(v_mT)\|< \overline{\alpha}\sqrt{\varepsilon}$. Hence, $h_i(v_mT)-h_j(v_mT)\rightarrow 0$, $1\leq i\neq j\leq N$, as $m\rightarrow \infty$ and $\varepsilon\rightarrow 0$. This together with Proposition 1, (\ref{eq12}), and  (\ref{eq5}) yields that $h_i(t)-h_j(t)\rightarrow 0$, $1\leq i\neq j\leq N$, as $t\rightarrow \infty$ and $\varepsilon\rightarrow 0$. Finally, according to Lemma 1, one can conclude that the multi-agent system (\ref{eq1}) achieves the practical output consensus specified by (\ref{eq41}). This completes the proof of Theorem 1. \IEEEQED

\end{document}